\documentclass[epj,final]{svjour}
\usepackage{graphics}
\begin{document}
\title{On confinement in a light-cone Hamiltonian for QCD} 
%
\dedication{Dedicated to Professor Heinz Horner, Heidelberg,
            at the occasion of his 60$^{th}$ birthday}
\author{H.C. Pauli\thanks{
        Hans-Christian Pauli,
        Max-Planck-Institut f\"ur Kernphysik,
	Postfach 10 39 80,
        D-69029 Heidelberg.}
}
 \offprints{\emph{H.C. Pauli}\\
  archive:  hep-th/9807xxx\hfill
  preprint: MPIH-V21-1998\\
  ftp: anonymous@goofy.mpi-hd.mpg.de/pub/pauli/confine/* 
  }
\institute{MPI Kernphysik, Heidelberg.
	   \email{pauli@mpi-hd.mpg.de}}
\date{Received: date / Revised version: 3 June 1998}
\abstract{
The canonical front form Hamiltonian for non-Abelian  SU(N)
gauge theory in 3+1 dimensions and in the light-cone gauge 
is mapped non-perturbatively on an effective  Hamiltonian 
which acts only in the Fock space of a quark and an antiquark.
Emphasis is put on the many-body
aspects of gauge field theory, and it is shown explicitly how
the higher Fock-space amplitudes can be retrieved 
self-consistently  from solutions in the $q\bar q$-space.
The approach is based on the novel method of iterated
resolvents and on discretized light-cone quantization 
driven to the continuum limit. It is free of the usual 
perturbative Tamm-Dancoff truncations in particle number 
and coupling constant and respects all symmetries of the
Lagrangian including covariance and gauge invariance. 
Approximations are done to the non-truncated formalism.
Together with vertex as opposed to
Fock-space regularization, the method allows to apply 
the renormalization programme non-perturbatively 
to a Hamiltonian. 
The conventional QCD scale is found arising
from regulating the transversal momenta.
It conspires with additional mass scales 
to produce possibly confinement.
} 
\maketitle

\section{Introduction} 
\label{sec:1} 

Over the past twenty years two fundamentally different pictures
of hadrons have developed. 
One, the constituent quark model is closely related to 
experimental observation and phenomenology, see for example  
\cite{goi85,dkm91,neu93,eiq94} and the references cited there.  
The other,  quantum chromodynamics (QCD) is based on a
covariant non-abelian quantum field theory. 
How can one reconcile such models with the need
to understand hadronic structure from a covariant theory?

Since ``the Hamiltonian is of central importance in usual 
quantum mechanics'' \cite {fey48}, an intuitive approach  
for solving relativistic bound-state problems would be 
to solve the gauge-fixed Hamiltonian eigenvalue problem. 
But the presence of the square root operator in the equal-time 
Hamiltonian approach presents severe mathematical difficulties. 
Even if these problems could be solved, the eigensolution is 
only determined in its rest system. 
Boosting such a wavefunction from the hadron's 
rest frame to a moving frame is as complex a problem as 
solving the bound state problem itself. 
This reflects the contemporary conviction that the concept of
a Hamiltonian is  old-fashioned  and littered with all kinds 
of almost intractable difficulties.  
Alternative procedures like those of
Schwinger and  Dyson or  of Bethe and Salpeter are not easy 
to cope with either in practice, see for example \cite{lsg91}.  

So, why does one not proceed with the only successful 
non-perturbative approach to gauge field theory, 
with lattice gauge theory \cite{fey48,wei94} 
and its recent quantitative predictions \cite{nrq95}?

It is the concept of a wavefunction which is so appealing
in an Hamiltonian approach, particularly if one chooses
the front form of Hamiltonian dynamics \cite{dir49}.
The light-cone wavefunctions encode the hadronic properties
in terms of their quark and gluon degrees of freedom, 
and thus all hadronic properties can be derived from them.  
One can compute electro-magnetic and weak form factors 
rather directly from an  overlap of light-cone 
wavefunctions \cite{leb80}, see also \cite{bpp97}, and the 
hadron and nuclear structure functions are 
comparatively simple probability distributions. 
Lattice gauge theory has still a long way to go to formulate
such dynamical aspects.

As recently reviewed \cite{bpp97}, one has several reasons 
\cite{wil89} why the front form of Hamiltonian 
dynamics is one of the very few candidates for getting wave functions, 
see also  \cite{brp91,gla95}. 
Particularly the simple vacuum and the simple boost properties
confront with the complicated vacuum and the complicated 
boosts in the conventional Hamiltonian theory.
Based on the recognition that
``the only truly successful approach to bound states in field theory
has been quantum electrodynamics (QED) with its combination
of non-relativistic quantum mechanics to handle bound states
and perturbation theory to handle relativistic effects'' \cite{wwh94},
Wilson and collaborators \cite{wwh94} have proposed a scheme
in which one presumes a potential for the bound-states
and handles the relativistic effects by structures imposed  
by the needs of renormalization.
This scheme continues to develop \cite{gla97}. The only available
numerical examples \cite{brp95,bpw97,jop96b} however
violate each and every symmetry of the Lagrangian.

The aim and motivation of the present work is similar to \cite{wwh94}.
One aims at finding an effective interaction between quarks
in analogy to the Coulomb potential as a crude and zero-order
approximation to a given Lagrangian which can be used 
as a starting point for more refined considerations
within a Hamiltonian approach.

But the problems one faces with a Hamiltonian 
are stupendous.
One has to deal with many difficult aspects of the theory 
simultaneously:
confinement, vacuum structure, spontaneous  breaking of 
chiral symmetry (for massless quarks), and  describing a
relativistic many-body system with unbounded particle number. 
The problem in non-Abelian gauge theory is compounded 
not only by the physics of confinement, 
but also by the fact that the wave function of a composite of
relativistic constituents has to describe systems of an arbitrary 
number of quanta with arbitrary momenta and helicities.  

For example, the vacuum in the front form is not truly 
simple \cite{hks92,vap92} but still simpler than in 
the instant form.
The problem can at least be formulated 
in terms of the `zero modes', the space-like constant
field components  defined in a finite spatial volume. 
The original conjecture that zero modes represent 
long range aspects and thus confinement \cite{kpp94},
however has not materialized \cite{kal96}.
Zero modes are important for a `theory of the vacuum',
but probably less for the massive part of the spectrum.
In the present work zero modes will therefore be suppressed
without any good argument other than simplicity. 
With similar arguments all aspects of chiral symmetry
(breaking) are disregarded. 

One addresses thus to find the bound-state structure
of the light-cone Hamiltonian in the light-cone gauge $A^+=0$.
This gauge is kind of natural to the light cone, and it is physical 
since the gluons have only two transverse degrees of freedom.  
Simple considerations show why other than in 1+1 dimensions  
one should not address to solve the full Hamiltonian \cite{bpp97} 
in physical space-time. In the first place one should address 
to develop a well-defined effective interaction, such one as
proposed for example by Tamm \cite{tam45} and independently by 
Dancoff \cite{dan50}, 
in their paradigmatic treatment of the Yukawa model.
But if one does so and adapts the Tamm-Dancoff
approach to the light-cone \cite{kpw92}, 
one finds no trace of a possible confinement in these `mesons'. 
This is rather disturbing since the same approach applied to 
positronium yields the Bohr aspects of the spectrum including the 
correct fine and hyperfine splittings.

The Tamm-Dancoff approach will be re-analyzed in Sect.~\ref{sec:3}. 
The problems are generic, 
since for the purpose of practical
calculations one must truncate at the lowest non-trivial
order of a perturbative series. 
If one tries to correct for that one gets into all kinds of difficulties
including the question how one should resume perturbative series to 
all orders without double counting.
The problem is a severe one since one is forced to mistreat 
precisely all those multi-particle configurations which one thinks 
are important for confinement in a field theory. Quite naturally, they
come in certain higher  powers of the coupling constant
and are thrown away when one truncates the 
perturbative series.
In a way that part of the problem is similar to throwing away 
many-body amplitudes in conventional many-body problems. 
The problem in field theory are accentuated by the fact
that the terms of higher order in the coupling constants 
are to be multiplied with very large numbers. 

The shortcomings of the Tamm-Dancoff approach are overcome 
here by the method of iterated resolvents 
\cite{pau93,pau96a,pau96}. 
This novel method is inspired by the Tamm-Dancoff approach and 
requires the development of a considerable formal apparatus. 
As to be shown, it allows for a systematic discussion of the 
many-body effects in a field theory without truncating the 
expressions at a finite order of perturbation theory, at any time.
With only a few well localizable assumptions one ends up 
with integral equations, whose structure is similar what was 
solved already in the past \cite{kpw92,pam95,trp96}. 
Most importantly one arrives at the conclusion that the 
effective interaction has essentially only two contributions:
a flavor-conserving and a flavor-changing part. 
In a diagrammatical representation both look like low order
perturbative graphs. In an approximative sense they even are
identical with perturbative graphs,  
with one important difference however: 
All genuine many-body effects reside in the effective coupling
constant and generate a dependence on the momentum transfer 
across the vertex. They generate new interactions and 
possibly generate confinement.

\section{The light-cone Hamiltonian (matrix)}
\label{sec:2} 

The advantages and challenges of quantizing field 
particularly gauge field theory `on the light cone' and of solving 
the bound state problem for quantum chromodynamics (QCD) 
have been reviewed recently \cite{bpp97}.
It should suffice therefore to recall here the most elementary 
facets and to shape some notational aspects in short.

\def\d{ T }  \def\v{ V }  \def\b{$\cdot$} \def\s{$\cdot$}   \def\f{$\cdot$}
\begin {table*}[t]
\begin {center}
\caption {\label {tab:3} \em
The Fock space and the Hamiltonian matrix $H^\prime =T+V$ 
for a meson at fixed value of $K=4$.~--- 
See discussion in the text.
The diagonal blocks are denoted by $T$. 
Most of the block matrices 
are zero matrices, marked by a dot ($\cdot$).
The block matrices marked by $V$ are potentiall non-zero 
due to the vertex interaction. 
}\vspace{1em}
\begin {tabular}  {||cc||cc||c|ccc|cccc|ccccc||}
\hline \hline 
     &   &        & N$_p$ & 
     2 & 2 & 3 & 4 & 3 & 4 & 5 & 6 & 4 & 5 & 6 & 7 & 8 
\\ 
   K & N$_p$ & Sector & n & 
     1 & 2 & 3 & 4 & 5 & 6 & 7 & 8 & 9 &10 &11 &12 &13 
\\ \hline \hline   
    1 &  2 & $ q\bar q\, $ &  1 & 
    \d &\s &\v &\f &\b &\f &\b &\b &\b &\b &\b &\b &\b 
\\ \hline
    2 &  2 & $ g\ g\ $ &  2 & 
    \s &\d &\v &\b &\v &\f &\b &\b &\f &\b &\b &\b &\b 
\\
    2 &  3 & $ q\bar q\, \ g $ &  3 & 
    \v &\v &\d &\v &\s &\v &\f &\b &\b &\f &\b &\b &\b 
\\
    2 &  4 & $ q\bar q\, q\bar q\, $ &  4 & 
    \f &\b &\v &\d &\b &\s &\v &\f &\b &\b &\f &\b &\b 
\\ \hline
    3 &  3 & $ g \ g \ g $ &  5 & 
    \b &\v &\s &\b &\d &\v &\b &\b &\v &\f &\b &\b &\b 
\\
    3 &  4 & $ q\bar q\, \ g \ g $ &  6 & 
    \f &\f &\v &\s &\v &\d &\v &\b &\s &\v &\f &\b &\b 
\\
    3 &  5 & $ q\bar q\, q\bar q\, \ g $ &  7 & 
    \b &\b &\f &\v &\b &\v &\d &\v &\b &\s &\v &\f &\b 
\\
    3 &  6 & $ q\bar q\, q\bar q\, q\bar q\, $ &  8 & 
    \b &\b &\b &\f &\b &\b &\v &\d &\b &\b &\s &\v &\f 
\\ \hline
    4 &  4 & $ g \ g \ g \ g $ &  9 & 
    \b &\f &\b &\b &\v &\s &\b &\b &\d &\v &\b &\b &\b 
\\
    4 &  5 & $ q\bar q\, \ g \ g \ g $ & 10 & 
    \b &\b &\f &\b &\f &\v &\s &\b &\v &\d &\v &\b &\b 
\\
    4 &  6 & $ q\bar q\, q\bar q\, \ g \ g $ & 11 & 
    \b &\b &\b &\f &\b &\f &\v &\s &\b &\v &\d &\v &\b 
\\
    4 &  7 & $ q\bar q\, q\bar q\, q\bar q\, \ g $ & 12 & 
    \b &\b &\b &\b &\b &\b &\f &\v &\b &\b &\v &\d &\v 
\\
    4 &  8 & $ q\bar q\, q\bar q\, q\bar q\, q\bar q\, $ & 13 & 
    \b &\b &\b &\b &\b &\b &\b &\f &\b &\b &\b &\v &\d 
\\ \hline\hline
\end {tabular}
\end {center}
\end {table*}

In canonical field theory the  four  commuting components of
the  energy-momentum four-vector $ P ^\nu$ are constants 
of the motion. 
In the front form of Hamiltonian dynamics \cite{dir49}
they are denoted by $ P ^\nu = (P  ^+, \vec P  _{\!\bot}, P  ^-)$, 
see also \cite{brp91,bpp97}. 
Its three spatial components 
$\vec P  _{\!\bot} =( P  ^1, P ^2)$  and $P  ^+$ 
do not depend on the interaction. 
The fourth component $P ^-= 2 P_+ $ depends on the
interaction  and is a complicated non-diagonal
operator. It propagates the system in light-cone 
time $ x^+ = x^0 + x^3$, {\it i.e.}  
$ i { \partial \over \partial x ^+ } \vert \Psi \rangle 
   = P _+ \vert \Psi \rangle $, 
and is the front-form Hamiltonian proper \cite{dir49}.
The contraction of  $P ^\mu$  is the operator of
invariant mass-squared, 
\begin {equation} 
      P ^\mu P _\mu  = P ^+ P ^- - \vec P  _{\!\bot} ^{\,2} 
      \qquad  \equiv H _{\rm LC} \equiv H
\ .\end {equation} 
It is a Lorentz scalar and  referred to
somewhat improperly, but conveniently, as the 
`light-cone Hamiltonian $ H _{\rm LC} $' 
\cite{brp91}, or shortly $ H $.
Its matrix elements (and eigenvalues) somewhat unusually 
carry the dimension of an invariant-mass-squared.

In this work one aims at a representation in which $H$ is diagonal, 
\begin {equation} 
      H \vert \Psi _i \rangle = E _i \vert \Psi _i \rangle
\,.\label{eq:2.2}\end {equation} 
One wants to calculate the bound-state spectrum $E _i $ 
together with the corresponding eigenfunctions $\Psi _i $ 
directly from the gauge field QCD-Lagrangian.

A convenient Hilbert space for the energy-momentum operators 
$P ^\nu$ is the {\em Fock space}.
The Fock space is the complete set of all possible {\em Fock states}
\begin {equation}
     \vert \Phi _n\rangle = 
     \ b ^\dagger _{q_1} b ^\dagger _{q_2} \dots 
     \ d ^\dagger _{q_1} d ^\dagger _{q_2} \dots 
     \ a ^\dagger _{q_1} a ^\dagger _{q_2} \dots
     \vert 0 \rangle
\,.\end {equation}
The creation and destruction operators
obey the familiar (anti-)commutation relations.
Each particle has four-momentum 
$ p^\mu = (p^+, \vec p _{\!\bot}, p^- )$ and
sits on its mass-shell $ p^\mu p_\mu = m^2$ 
with (light-cone) energy 
$p^- = (m^2 + \vec p ^{\,2}_{\!\bot}) / p^+ $.
Each quark state is specified by the six quantum numbers 
$ q = (p^+, \vec p_{\!\bot}, \lambda; c, f) $, {\it i.e.}
the three spatial momenta, helicity, color and flavor,
respectively. Correspondingly, gluons are specified by 
five quantum numbers $q = (p^+, \vec p_{\!\bot}, \lambda;a) $ 
with the glue index $a$. 
The three spatial components $\vec P  _{\!\bot}$  and $P  ^+$ 
are diagonal operators in Fock space representation with
eigenvalues
\begin {equation}
     P ^+ = \sum _{n} p^+ _{n} 
     \ , \quad {\rm and} \quad
     \vec P  _{\!\bot} = \sum _{n} (\vec p _{\!\bot} )_{n} 
\ . \label{eq:2.1}\end {equation}
The matrix elements of $P^-$ (or of $H$) in Fock-space 
representation are given elsewhere \cite{bpp97}.

The eigenvalue equation in Eq.(\ref{eq:2.2}) stands for an infinite set of 
coupled integral equations which are extremely difficult to handle, 
see \cite{bpp97}. It is useful to convert it to the much more transparent
case of  a {\em finite set} of coupled matrix equations, namely 
by the technical trick of imposing periodic  boundary conditions
(DLCQ, \cite{pab85a}). 
Introducing a box size $L$ as a finite and additional length 
parameter, however, can be at most an intermediate step. 
Latest at the end of the calculations, it must be removed 
by a limiting procedure  like
$L\longrightarrow \infty$, $K\longrightarrow \infty$, but $K/L$ finite,
since only the continuum can be the full, covariant theory.

Why is this set finite?~---
The longitudinal light-cone momentum $p^+$ is a positive number.
For periodic boundary conditions the lowest possible value is
$(p^+)_{\rm min}= \pi/L$.
(Zero modes with $p^+=0$ are disregarded here, as mentioned.)
Consequently, any total momentum $P^+= K\pi/L$
can be distributed over at most $K$ bosons, or over
$K$ fermion pairs since these are subjected to 
anti-periodic boundary conditions.
As illustrated in Table~\ref{tab:3} for the Fock space of a meson,
the {\em harmonic resolution} $K$ \cite{pab85a} governs 
the number of Fock space sectors.

The lowest possible value $K=1$ allows  only for one 
Fock-space sector with a single $q\bar q$-pair 
(a single gluon cannot be in a color singlet). 
For  $K=2$, the Fock space  contains in addition two gluons, 
a $q\bar q$-pair plus a gluon, and two $q\bar q$-pairs. 
For $K=4$ the Fock space contains at most 8 particles. 
One can label the Fock space sectors according to the
the quark-gluon content, or one can enumerate them, 
which is less transparent but more simple.
In Table~\ref{tab:3} the Fock-space sectors 
for $K\leq 4$ are enumerated $n=1,\dots,13$.
With increasing $K$ more Fock-space sectors are added.  
Their total number grows like $N_K =  (K+1)(K+2)/2- 2 $. 

Once the Fock space is defined, one can calculate the 
Hamiltonian matrix. 
In  \cite{bpp97} $H$ is divided into three structurally different parts:
\begin{equation}
         H = T + V + W
\,.\end{equation}
The kinetic energy  $T$ is independent of the 
coupling constant $g$. 
It is diagonal in Fock-space representation, 
and contributes only to the diagonal blocks in Table~\ref{tab:3}. 
These diagonal blocks carry only diagonal matrix elements 
which are given by Eq.(\ref{eq:m6}).~---
The {\em vertex interaction} $V$ is the relativistic interaction
{\it per se} and linear in $g$. 
The odd number of creation and destruction operators 
prevents diagonal matrix elements. 
Potentially, the vertex interaction has non-zero matrix elements 
only between Fock-space sectors whose particle numbers 
differ by 3 or by 1.
Those differing by 3 are the typical vacuum fluctuation diagrams
in which a fermion pair is created simultaneously with a gluon. 
On the light-cone, they vanish kinematically because
of longitudinal momentum conservation:
{\em The vacuum does not fluctuate},
see for example \cite{bpp97}.
In consequence, the vertex interaction on the light cone 
can change the particle number only by 1.
The corresponding blocks are marked by a $V$ in
Table~\ref{tab:3}.~---
The {\em instantaneous interaction} $W$ arises as a 
consequence of working in the light-cone gauge.
Potentially, it has non-zero matrix elements between
Fock space sectors whose particle number differs by 0 or 2.
In the following discussion $W$ will be left out  
because of a much more transparent formalism.  
The impact of $W$ will be restored by a simple trick 
towards the end of Section~\ref{sec:3}. 

Table~\ref{tab:3} highlights some of the main issues of a
Hamiltonian approach.
It demonstrates that the Hamiltonian matrix is very sparse:
most of the block matrices between the sectors are plain 
zero matrices.
Very much like in the non-relativistic case with its typical pair-interactions,
the Hamiltonian cannot connect all Fock-space sectors due to
the selection rules imposed by the creation and destruction operators. 
Depending on the arrangement of the sectors,
the Hamiltonian matrix has a (tri-diagonal) band structure. 
The table illustrates further that 
the  Hamiltonian `on the light cone' is separable into a kinetic part 
and the interaction, again like in conventional non-relativistic 
many-body problems. 
Therefore it should be approachable like that, namely by 
introducing some complete and denumerable set of functions
$\vert n\rangle$, in terms of which one can calculate the
`Hamiltonian matrix' $\langle n\vert H\vert m\rangle$.
Its diagonalization, 
\begin{equation}
        \sum_{n,n^\prime} \langle \Psi_{j}\vert n^\prime\rangle
        \langle n^\prime\vert H\vert n\rangle 
        \,\langle n\vert\Psi_{i}\rangle = 
        E_{i} \delta_{ij}
\,, \end {equation} 
is equivalent to solving Eq.(\ref{eq:2.2}).
The columns of the unitary transformation matrix 
$\langle n\vert\Psi_{i}\rangle $ are 
the `Fock-space amplitudes' or `wave functions'.
The eigenstates are complicated superpositions of  them, {\it i.e.} 
$
   \vert\Psi_{i}\rangle =
   \sum_n \vert n\rangle\,\langle n\vert\Psi_{i}\rangle
$.

But the analogue with non-relativistic Hamiltonian is superficial.
There, the diagonal blocks $\langle n\vert H\vert n\rangle$ have
off-diagonal matrix elements (potential energies) and one can
approach the problem by successive truncation.
In (gauge) field theory, the diagonal blocks contain only the
(diagonal) kinetic energies.
Possible binding effects most come from the virtual scattering 
to the higher sectors.
Truncating the matrix as done so often in many-body theory 
prevents such virtual scattering and 
potentially violates gauge invariance. Moreover, since 
the theory exposes divergies which can be controlled
by a cut-off or regulator scale $\Lambda$,
all eigenvalues will depend on it: 
$E_i = E_i(g,m_f;\Lambda)$.
Since this is unphysical, one must require for all of them
that they are independent of $\Lambda$, {\it i.e.}
\begin {equation}
        {d\over d\Lambda}E_i(g,m_f;\Lambda) = 0
\,.\label{eq:i69}\end {equation} 
Non-perturbative renormalization has not yet been applied 
to a Hamiltonian, in practice.

The bottle neck of any Hamiltonian approach 
as illustrated in Table~\ref{tab:3} as well: 
The dimension of the Hamiltonian matrix  increases exponentially. 
To give an example, suppose the regularization procedure 
allows for 10 discrete
momentum states in each direction, {\it i.e.} in the one 
longitudinal and the two transversal directions of $\vec k_{\!\perp}$. 
A particle has then roughly $10^3$ degrees of freedom.
A Fock-space sector with $n$ constituent particles
has thus $10^{n-1}$ different Fock states, since they
are subject to the constraints, Eq.(\ref{eq:2.1}).
Sector 13 alone with its  8 particles has thus the 
dimension of  about $10^{21}$ Fock states.
The Hamiltonian method applied to gauge theory therefore 
faces a formidable matrix  diagonalization problem.
Sooner or later, the matrix dimension exceeds
imagination, and other than in 1+1 dimensions 
one has to develop new tools by matter of principle.

Aiming at an effective interaction between a quark and an antiquark,
different novel methods have been proposed recently.
Glazek and Wilson \cite{wwh94} propose to pre-diagonalize 
the Hamiltonian approximatively but analytically
such that its band width becomes narrower.  
The work on an iterative procedure is still going on \cite{gla97}. 
First applications to heavy mesons \cite{bpw97} have been 
carried out, but it is unclear how one can correct for the admitted
violation of every possible symmetry.
Wegner \cite{weg94} has proposed an analytic unitary transform 
which leads to Hamiltonian flow equations.
Applications to realistic models \cite{lew96} are promising.
Preliminary studies on QED \cite{guw97} are available.
In parallel and partially prior to these works,
the method of iterated resolvents 
\cite{pau93,pau96a,pau96}
has been proposed.
Its consequences are worked out in the sequel.

\section{Fock-space and vertex regularization} 
\label{sec:a} 

Before proceeding with the theory of effective interactions, 
the regularization of the theory is specified next in order 
to face a well defined and finite Hamiltonian matrix.

As mentioned, 
the finite number of Fock-space sectors is a consequence 
of the positivity of the longitudinal light-cone momentum $p^+$.
The transversal momenta $\vec p_{\!\bot}$ can take
either sign, and the number of Fock states within each sector
can be arbitrarily large.
In order to face a finite dimensional Hamiltonian matrix
one must have a finite number of Fock states, and this is 
achieved by {\em Fock space regularization}:
Following Lepage and Brodsky \cite{leb80}, a Fock state 
with $n$ particles is included
only if its {\em kinetic energy $T$},
\begin{equation}
         T_n = 
         \left(p_1+p_2+\dots p_n\right)^2 =         
         \sum_{\nu=1}^{n} 
         \left( {m^2+ \vec k_{\!\perp}^{\,2}\over x} \right)_{\nu}
\,,\label{eq:m6}\end{equation}
does not exceed a certain threshold.
$T$ can be interpreted as the free invariant mass (squared)
of the Fock state.
The lowest possible value of  $T_n$ is taken when 
all particles are at rest relative to each other, {\it i.e.}
$\left(T_n\right)_{min}=\left(m_1+m_2+\dots m_n\right)^2$. 
This frozen invariant mass should be removed from the cut-off 
\begin{equation}
         \sum_{\nu=1}^{n} 
         \left( {m^2+ \vec k_{\!\perp}^{\,2}\over x} \right)_{\nu} -
         \left( m_1+m_2+\dots m_n\right)^2
         \leq \Lambda_{\rm FS} ^2 
\,.\end{equation}
Since $x$ and $\vec k_{\!\perp}$ are the usual 
{\em momentum fractions} and 
{\em intrinsic transversal momenta}, respectively,
the regularization is {\em frame-independent} \cite{bpp97}.
The mass scale $\Lambda_{\rm FS} $ is a Lorentz scalar 
and one of the parameters of the theory.

However, it was not realized in the past \cite{brp91}, that 
Fock-space regularization is almost irrelevant in the 
continuum theory. {\em Vertex regularization} is a better alternative.~--
At each vertex,  a particle with four-momentum 
$p^\mu$ is scattered into two particles with respective 
four-momentum $p_1^\mu$ and $p_2^\mu$. 
Parameterizing the momenta as 
$p^\mu=(p^+,\vec p_{\!\perp},p^-)$,  
$p_1^\mu=(zp^+,z\vec p_{\!\perp}+\vec l_{\!\perp},p_1^-)$ and
$p_2^\mu=\left(\left(1-z\right)p^+,
  \left(1-z\right)\vec p_{\!\perp}-\vec l_{\!\perp},p_2^-\right)$,
the (vertex) matrix element \cite{brp91} is proportional to
$\vec l_{\!\perp}^{\ 2}/z$.  It tends to diverge for
$l_{\!\perp}\rightarrow\infty$ and/or $z\rightarrow 0$.
In order to avoid potential singularities 
{\em one can regulate the interaction}
by setting the matrix element to zero if the off-shell mass 
$(p_1+p_2)^2$ exceeds a certain scale  $\Lambda$.
The condition
\begin{equation} 
      (p_1+p_2)^2-( m_1+m_2)^2 \le \Lambda^2 
\,\label{eq:i1}\end{equation} 
will be referred to as {\em vertex regularization}. 
The off-shell mass 
\begin{eqnarray} 
\lefteqn{ 
   (p_1+p_2)^2-( m_1+m_2)^2=
}\nonumber\\ & &
   {1\over z(1-z)}\left( \vec l_{\!\perp}^{\ 2}+
   \left(m_1+m_2\right)^2 \left( z-\overline z \right)^2 \right) 
\,,\label{eq:i2}\end{eqnarray} 
governs how much the particles can go off their 
equilibrium values ${l} _{\!\perp}=0$ and 
$\overline z= m_1/(m_1+m_2)$.
For $\Lambda\rightarrow 0$, the phase space is reduced to a point, 
and consequently the interaction is reduced to zero: 
The Hamiltonian matrix (or the integral equation) is diagonal. 
Irrespective of the matrix dimension governed by 
$\Lambda_{\rm FS}$, the spectrum of the interacting theory
is identical with the free theory.

Vertex regularization Eq.(\ref{eq:i1}) is realized by 
the cut-off function $\Theta$,  {\it i.e.}
\begin {equation} 
       \Theta(z,l_{\!\perp}) = 
       \cases{
       1, &for $\ 0 \leq l_{\!\perp}^2 \leq  l^2_{\Lambda} (z)$,
            $\ \epsilon_l\leq z \leq 1 - \epsilon_u\ $, \cr
       0, &otherwise.\cr}
\label{eq:i3}\end{equation}
The limiting momentum $l_\Lambda^2(z)$ 
describes a semi-circle in the appropriate units, 
\begin{eqnarray} 
      l_\Lambda^2(z)&=&\big(\Lambda^2+(m_1+m_2)^2\big)
      \bigg[\bigg({c\over2}\bigg)^2 -\bigg(z-{b\over2}\bigg)^2\bigg]
\nonumber\\ \quad{\rm with}\quad
       b&=&{\Lambda^2+2m_1(m_1+m_2)\over
       \Lambda^2+(m_1+m_2)^2} 
\nonumber\\ \quad{\rm and}\quad
       c^2&=&{\Lambda^2\over\Lambda^2+(m_1+m_2)^2}
       {\Lambda^2+4m_1m_2\over\Lambda^2+(m_1+m_2)^2}
\,.\label{eq:m94}\end{eqnarray} 
Note that the semi-circle intersects the $z$-axis at 
$z_1=\epsilon_l$ and $z_2=1 - \epsilon_u$, with 
\begin {equation} 
       \epsilon_l={b-c\over2} \quad{\rm and}\quad
       \epsilon_u=1-{b+c\over2} 
\,.\end{equation}
For sufficiently large $\Lambda$
and equal masses ($m_1=m_2=m$), the limits become
$\epsilon_l=\epsilon_u= (m/\Lambda)^2$.

The scale parameter $\Lambda$ 
regulates potential transversal divergences 
($l_{\!\perp}\rightarrow\infty$), 
while potential longitudinal singularities ($z\rightarrow 0$)
are regulated by the single particle mass $m$.
If all particles are endorsed with a
small additional regulator mass according to
\begin{equation} 
   m^2 \longrightarrow m^2 + m_{reg}^2 
\,,\label{eq:m15}\end{equation}
one ensures that the point $z=0$ is never inside the 
semi-circle given by Eq.(\ref{eq:m94}), even not for
the massless gluons, or for the quarks in
the limit $m_f\rightarrow 0$. 
Vertex regularization regulates then all divergences 
on the light cone. 

\section {The method of iterated resolvents}
\label{sec:3}

Effective interactions are a well known tool in  many-body 
physics \cite{mof50}. In field theory
the method is known as the Tamm-Dancoff-approach, 
applied first by Tamm \cite{tam45} and by Dancoff \cite{dan50}. 
Let us review it in short.

As explained above, the Hamiltonian
matrix can be understood as a matrix of block matrices,
whose rows and columns are enumerated by
$i=1,2,\dots N$ in accord with the  Fock-space sectors
in  Table~\ref{tab:3}.
Correspondingly, one can rewrite Eq.(\ref{eq:2.2})  
as a block matrix equation:
\begin {equation} 
      \sum _{j=1} ^{N} 
      \ \langle i \vert H \vert j \rangle 
      \ \langle i \vert \Psi\rangle 
      = E\ \langle i \vert \Psi\rangle 
\, \qquad {\rm for\ all\ } i = 1,2,\dots,N 
\ .\label {eq:319}\end {equation} 
The rows and columns of the matrix can always be split
into two parts. One speaks of  the $ P $-space  
$P = \sum _{j=1} ^n  \vert j \rangle\langle j \vert $
with $1<n<N$, and of the rest, 
the $Q$-space $Q\equiv 1-P$.
Eq.(\ref{eq:319}) can then be rewritten conveniently 
as a $2\times2$ block matrix equation
\begin {eqnarray} 
   \langle P \vert H \vert P \rangle\ \langle P \vert\Psi\rangle 
 + \langle P \vert H \vert Q \rangle\ \langle Q \vert\Psi\rangle 
 &=& E \:\langle P \vert \Psi \rangle 
 ,  \label{eq:321} \\ 
   \langle Q \vert H \vert P \rangle\ \langle P \vert\Psi\rangle 
 + \langle Q \vert H \vert Q \rangle\ \langle Q \vert\Psi\rangle 
 &=& E \:\langle Q \vert \Psi \rangle 
. \label{eq:322}\end {eqnarray}
One rewrites the second equation as
$    \langle Q \vert E  -  H \vert Q \rangle 
    \ \langle Q \vert \Psi \rangle 
  =   \langle Q \vert H \vert P \rangle 
    \ \langle P \vert\Psi\rangle $,
and observes that the quadratic matrix 
$ \langle Q\vert E -  H \vert Q \rangle $ could be inverted 
to express the Q-space  
in terms of the $ P $-space wavefunction. 
But here is a problem:   
The eigenvalue $ E $ is unknown at this point. 
One therefore solves first an other problem: One introduces
{\em the starting point energy} $\omega$ as a redundant
parameter  at disposal, and 
{\em defines the $Q$-space resolvent} 
as the inverse of the block matrix 
$\langle Q \vert\omega- H \vert Q \rangle$.
The $Q$-space function becomes then
\begin{eqnarray} 
   \langle Q \vert \Psi (\omega)\rangle    = & &
   G _ Q (\omega) \langle Q \vert H \vert P \rangle 
   \,\langle P \vert\Psi\rangle 
,\nonumber\\ & &
   G _ Q (\omega) =  
   {1\over\langle Q \vert\omega- H \vert Q \rangle} 
.\label{eq:332}\end{eqnarray}
If there is no danger of confusion, the argument in the resolvents 
$G (\omega)$ will often be dropped in the sequel.
Inserting it into Eq.(\ref{eq:321}) defines the effective 
Hamiltonian
\begin{equation} 
      H _{\rm eff} (\omega) =  H +
      H \vert Q \rangle\,G_Q(\omega)\,\langle Q\vert H 
\,.\label{eq:340}\end{equation} 
By construction it acts only in the $P$-space
\begin{equation} 
      H _{\rm eff} (\omega ) 
      \vert P\rangle\,\langle P \vert\Psi_k(\omega)\rangle =
      E _k (\omega )\,\vert\Psi _k (\omega ) \rangle 
\,.\label{eq:345}\end{equation} 
In addition to the original Hamiltonian in the $P$-space, 
the effective Hamiltonian 
acquires a piece where the {\em system is scattered virtually into the 
higher sectors} represented by the $Q$-space, 
propagating there ($G_Q$) by impact of the true interaction, 
and scattered back into the $ P $-space. 
Every value of $\omega$ defines a different Hamiltonian 
and a different spectrum. 
Varying $\omega$ one generates a set of  
{\em energy functions} $ E _k(\omega) $. 
Whenever one finds a solution to the {\em fix-point equation} 
\begin {equation}
E _k (\omega ) = \omega 
, \label {eq:350} \end {equation}
one has found one of the true eigenvalues and
eigenfunctions of $H$,  by construction. 

One should emphasize that one can find all eigenvalues of 
of the full Hamiltonian $H$,  
irrespective of how small one chooses the $ P $-space. 
Explicit examples for that can be found in \cite{pau96}. 
It looks as if one has mapped a difficult problem, the 
diagonalization of a huge matrix onto a simpler problem,  
the diagonalization of a much smaller matrix. 
But this true only in a restricted sense. 
One has to invert a matrix. The numerical inversion of a
matrix takes  about the same effort as its diagonalization. 
In addition,  one has to vary $\omega$ and solve the fix-point 
equation (\ref{eq:350}).  The numerical work is thus rather 
larger than smaller as compared to a direct diagonalization. 

The advantage of working with an effective interaction is 
of analytical nature to the extent that resolvents can be
approximated systematically.  The two resolvents
\begin{eqnarray} 
     G _Q (\omega) &=&  {1\over \langle Q \vert 
           \omega - T - U  \vert Q \rangle} 
     ,\quad{\rm and}\quad 
\nonumber\\ 
     G _0  (\omega) &=& {1\over \langle Q \vert 
           \omega - T \vert Q \rangle} 
, \label{eq:352}\end{eqnarray} 
defined once with and once without the non-diagonal 
interaction $U$, are identically related by 
$G_Q=G_0 +G_0 U G_Q $,
or by the infinite series of perturbation theory.
\begin{equation}  
    G _Q =
    G _0 + G_0 U G _0 +
    G_0 U G_0 U G _0 + \dots 
.\label{eq:m24}\end{equation} 
The point is that the kinetic energy $T$ is
a diagonal operator which can be trivially inverted to get 
the unperturbed resolvent $G_0$. 

In practice, Tamm and Dancoff \cite{tam45,dan50} have restricted 
themselves to the first non-trivial order, and also the recent 
applications to the light-cone Hamiltonian have
not gone beyond that \cite{kpw92,trp96}. 
This is unsatisfactory, since it destroys  
Lorentz and gauge invariance. 
Even worse, if one identifies $\omega$ with the eigenvalue 
(as one should), the effective interaction develops a 
non-integrable  singularity \cite{tam45,dan50}. 
In the front form work \cite{kpw92,trp96} it was argued why
the so called $\omega=\omega^\star$ trick removes this 
singularity and approximatively restores gauge invariance.
In the essence, one replaces the eigenvalue $\omega$ by 
the average kinetic energy in the $P$-space.

The Tamm-Dancoff approach can be interpreted as the reduction 
of a block matrix dimension from 2 to 1, particularly the step from 
Eqs.(\ref{eq:321},\ref{eq:322})  to Eq.(\ref{eq:345}). 
But there is no need for identifying the 
$P$-space with the lowest sector. In the sequel one chooses
the last sector as the $Q$-space: 
The same steps as above reduce then the block matrix dimension 
from $N$ to $N-1$. The effective interaction acts in the now 
smaller space. This procedure can be iterated. 
The disadvantage is that one deals with `resolvents of resolvents', 
or with {\em iterated resolvents}. The advantage is that 
the zero block matrices simplify the algorithm considerably.
In the Tamm-Dancoff approach they cannot be made use of.
Ultimately, one arrives at block matrix dimension 1 where the 
procedure stops: The effective interaction in the Fock-space
sector with only one quark and one antiquark is defined
unambiguously \cite{pau96}.

Suppose, in the course of this reduction, one has 
arrived at block matrix dimension $n$, with $1\leq n\leq N$.  
Denote the corresponding effective interaction  $H_n (\omega)$. 
Since one starts from the full  Hamiltonian in the last 
sector $N$, one has to convene that $H_{N}\equiv H$. 
The eigenvalue problem corresponding to  
Eq.(\ref{eq:345}) reads then
\begin{equation} 
   \sum _{j=1} ^{n} \langle i \vert H _n (\omega)\vert j \rangle 
                    \langle j \vert\Psi  (\omega)\rangle 
   =  E (\omega)\ \langle i \vert\Psi (\omega)\rangle 
, \label{eq:406}\end {equation}
for $ i=1,2,\dots,n $.
Observe that $i$ and $j$ refer here to sector numbers.  
Now,  in analogy to Eq.(\ref{eq:332}), define 
\begin{eqnarray} 
   \langle n \vert \Psi (\omega)\rangle   
   = & & G _ n (\omega) 
   \sum _{j=1} ^{n-1} \langle n \vert H _n (\omega)\vert j \rangle 
   \ \langle j \vert \Psi (\omega) \rangle 
,\nonumber\\ 
    & & G _ n (\omega)   
    =   {1\over \langle n \vert\omega- H_n (\omega)\vert n \rangle} 
. \label{eq:410}\end{eqnarray}
The effective interaction 
in the  ($n -1$)-space becomes then 
\begin {equation}  
       H _{n -1} (\omega) =  H _n (\omega)
  +  H _n(\omega) G _ n  (\omega) H _n (\omega)
\ .\label {eq:414} \end {equation}
This holds for every block matrix  element 
$\langle i \vert H _{n-1}(\omega)\vert j \rangle$.  
To get the corresponding eigenvalue equation one
substitutes $n$ by $n-1$ everywhere in Eq.(\ref{eq:406}). 
Everything proceeds like in section~\ref{sec:2}, 
including the fixed point equation  $ E  (\omega ) = \omega $.
But one has achieved much more: Eq.(\ref{eq:414}) is a 
{\em recursion relation} which holds for all $1<n<N$!

The method of iterated resolvents \cite{pau93,pau96a,pau96}
is applicable to any many-body theory.
But due to the Fock-space structure as displayed in Table~\ref{tab:3}
it is particularly easy to apply it to gauge theory.
Let us demonstrate that in a stepwise procedure. 
For $K=1$ the Fock space has only one Fock-space sector,
and the effective Hamiltonian is simply the kinetic energy,
{\it i.e.} $H_{1}=T_{1}$. 
For $K=2$ one finds in Table~\ref{tab:3} that  the Fock space 
has $N_K=4$ sectors. 
By definition, the last sector contains only the diagonal kinetic 
energy, thus  $H_{4}=T_{4}$. 
Its resolvent $G_{4}(\omega)$ is calculated trivially. 
Then $H_{3}$ can be constructed, and thus $G_{3}(\omega)$
by a matrix inversion, followed by $H_{2}$ and finally $H_{1}$.
Dropping $\omega$ in the notation of the propagators,
for simplicity, one gets consecutively
\begin {eqnarray} 
     H _4 &=&  T_4 
\,,\label{eq:m33}\\  
     H _3 &=&  T_3 + V G _4 V 
\,,\\  
     H _2 &=&  T_2 + V G _3 V  
\,,\\  
     H _1 &=&  T_1 + V G _3 V + V G _3 V  G _2 V G _3 V 
\,.\label{eq:m36}\end {eqnarray}
The effective Hamiltonian $H _1$ with its two terms is
remarkably simple particularly when compared with the 
infinite series of the Tamm-Dancoff approach in Eq.(\ref{eq:m24}).
The method of iterated resolvents makes efficient
use of the sparseness of the gauge field Hamiltonian and
its many zero block matrices.

To get the effective Hamiltonian(s) for harmonic resolution(s)
$K=3,4,\dots$ is not repeated here explicitly.
Important is the general feature that the effective 
sector Hamiltonians are separable in the kinetic energies $T$ 
and the effective interactions $U(\omega)$,
\begin{equation}
       H _n(\omega) = T_n + U_n(\omega) 
\,.\end{equation}
Important is also that the effective Hamiltonians in the lower 
sectors become independent of $K$. 
The transition to the {\em continuum limit} $K\rightarrow\infty$ 
is then rather trivial and will hence forward be assumed. 
For the sake of future applications the effective interaction 
was calculated for the first 12 sectors.
Grouping them  in a different order, 
one finds with the short-hand notation of Table~\ref{tab:3} 
\begin{eqnarray} 
     U_1 
&=&  V G _3 V + V G _3 V  G _2 V G _3 V 
,\label{eq:610}\\  
     U_3 
&=& V G _6 V + V G _6 V  G _5 V G _6 V + V G _4 V 
,\label{eq:620}\\  
     U_6 
&=&  V G _ {10} V+ V G _{10} V  G _9 V G _ {10} V + V G _7 V 
,\label{eq:621}\\
     U_{10} 
&=&  V G _{15} V + V G _{15} V  G _{14} V G _{15}V +V G _{11} V  
,\label{eq:622} \end {eqnarray}
for the sectors with one $q\bar q$-pair, {\it i.e.} for
$q\bar q$,
$q\bar q\,g$,
$q\bar q\,gg$, and
$q\bar q\,ggg$, respectively. 
The effective interactions in the sectors with two 
$q\bar q $-pairs are:
\begin {eqnarray} 
    U_4 
&=& V G _7 V + V G _7 V  G _6 V G _7 V 
,\\  
    U_7 
&=& V G _ {11} V+ V G _{11} V  G _{10} V G _ {11} V + V G _8 V 
,\\
    U_{11} 
&=& V G _{16} V + V G _{16} V  G _{15} V G _{16}V  + V G _{12} V 
,\end {eqnarray}
for
$q\bar q q\bar q$,
$q\bar q q\bar q\, g$, and
$q\bar q q\bar q\, gg$.
Those with three $q\bar q $-pairs have 
\begin {eqnarray} 
    U_8 
&=& V G _{12} V + V G _{12} V  G _{11} V G _{12}V 
,\\
    U_{12}
&=& V G _{17} V + V G _{17} V  G _{16} V G _{12}V + V G _{13} V 
,\end {eqnarray}
for 
$q\bar q q\bar q q\bar q$ and 
$q\bar q q\bar q q\bar q\, g$.
The structure of the effective interaction in the pure glue sectors
is different:
\begin {eqnarray} 
     U_{gg} = U_2 
&=& V G _3 V + V G _5 V 
\,,\label{eq:643}\\  
     U_{gg\, g} = U_5 
&=& V G _ {6} V + V G _{9} V 
\,,\\
     U_{gg\, gg} = U_9 
&=& V G _{10} V + V G _{14} V  
\,.\label{eq:645}\end {eqnarray}
One notes finally that Eqs. (\ref{eq:m33})-(\ref{eq:m36}) can be 
reconstructed easily from the above relations 
by setting formally to zero all propagators $G _{n} $ with $n\ge 5$.

Due to the peculiar structure of the gauge field Hamiltonian 
in Table~\ref{tab:3}, the vertex interaction appears in the effective
interactions only in even pairs, typically in the combination  $VGV$. 
It is then plausible that  by simply substituting 
\begin{equation}
       VGV\rightarrow W+VGV
\,\end{equation}
one restores the instantaneous interaction $W$ which was omitted 
thus far. This rule was checked explicitly in rather laborious 
calculations in \cite{pau96}. 
It finds its counterpart in the rules for perturbative diagrams
\cite{leb80}, where every intrinsic `dynamic' line must be 
supplemented with the corresponding `instantaneous' line 
order by order in perturbation theory.

\begin{figure} 
  \resizebox{0.48\textwidth}{!}{%
  \includegraphics{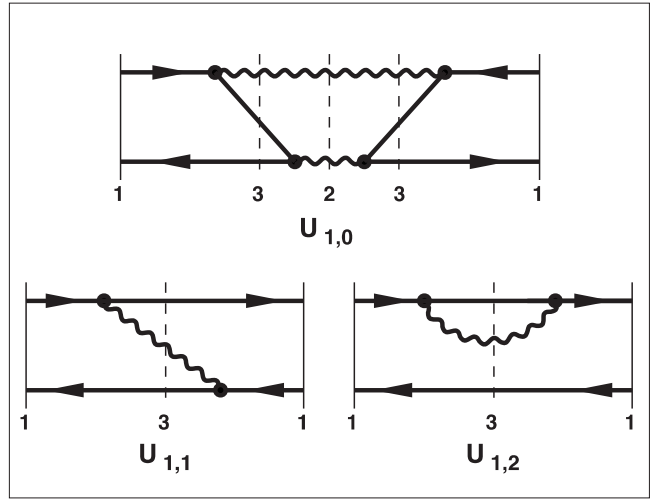} 
} \caption{
  The three graphs of the effective interaction in the 
  $q\bar q$-space.~---
  The lower two graphs correspond to the chain $U=VG_3V$,
    the upper corresponds to $U_a=VG_3V G_2V G_3V$. 
   Propagator boxes are represented by vertical dashed lines, 
   with the subscript `$n$'  refering to the sector numbers. 
} \label{fig:6_1}
\end{figure}

The most important result of this section is that gauge theory 
particularly QCD has only two structurally different contributions
to the effective interaction in the $q\bar q$-space,
see Eq.(\ref{eq:610}).
The effective one-gluon exchange
\begin{equation}
             U = V G _3 V
\end{equation}
conserves the flavor along the quark line and describes all fine 
and hyperfine interactions.
As illustrated in Fig.~\ref{fig:6_1} the vertex interaction $V$ 
creates a gluon and scatters the system virtually into the 
$q\bar q\,g$-space. 
As indicated in the figure by the vertical line with subscript `3', 
the three particles propagate there under impact of the full 
Hamiltonian before the gluon is  absorbed. 
The gluon can be absorbed either by the antiquark or by the 
quark. If it is absorbed by the quark, it contributes to the
effective quark mass $\overline  m$.  
The second term in Eq.(\ref{eq:610}), 
the effective two-gluon-annihilation interaction,
\begin{equation}
             U_a = V G _3 V G _2 V G _3 V 
\,,\end{equation}
is represented by the graph 
$U_{1,0}$ in Fig.~\ref{fig:6_1}. 
The virtual  annihilation of the $q\bar q$-pair into two gluons 
can generate an interaction between different quark flavors. 
As a net result the effective interaction scatters a quark with 
helicity $\lambda_q$ and four-momentum 
$p = (xP^+, x\vec P_{\!\perp} + \vec k_{\!\perp}, p^-)$
into a state with $\lambda_q^\prime$ and four-momentum 
${p^\prime} = (x^\prime P^+, x^\prime\vec P_{\!\perp} 
+ \vec k_{\!\perp}^\prime, {p^\prime}^-)$, 
and correspondingly the antiquark.
In the {\em continuum limit}, the resolvents are replaced by 
propagators and the eigenvalue problem 
$H _{\rm eff}\vert \psi\rangle=M^2\vert \psi\rangle$ 
becomes again an integral equation, but a very simple one
in only three continuous variables ($x,\vec k_{\!\perp}$).
Its structure is rather transparent:
\begin{eqnarray} 
\lefteqn{ 
    M_i^2\langle x,\vec k_{\!\perp}; \lambda_{q},
    \lambda_{\bar q}  \vert \psi _i\rangle 
}\nonumber\\ & & =
    \left[ {\overline m^2_{q} + \vec k_{\!\perp}^2 \over x } +
    {\overline m^2_{\bar  q} + \vec k_{\!\perp}^2 \over 1-x }\right]
    \langle x,\vec k_{\!\perp}; \lambda_{q},
    \lambda_{\bar q}  \vert \psi _i\rangle 
\nonumber\\ & &
    +\sum _{ \lambda_q^\prime,\lambda_{\bar q}^\prime}
    \!\int\!dx^\prime d^2 \vec k_{\!\perp}^\prime\,
    \,\Theta(x^\prime,k_{\perp}^\prime)     
    \,\langle x,\vec k_{\!\perp}; \lambda_{q}, \lambda_{\bar q}
    \vert V G _3 V +
\nonumber\\ & &
    V G _3 V G _2 V G _3 V 
    \vert x^\prime,\vec k_{\!\perp}^\prime; 
    \lambda_{q}^\prime, \lambda_{\bar q}^\prime\rangle\,
    \langle x^\prime,\vec k_{\!\perp}^\prime; 
    \lambda_{q}^\prime,\lambda_{\bar q}^\prime  
    \vert \psi _i\rangle 
.\label{eq:445}\end {eqnarray}
The domain of integration is set by the sharp cut-off function 
$\Theta$ given in Eq.(\ref{eq:i3}).
The eigenvalues refer to the invariant mass $M_i$ of a physical 
state. The wavefunction 
$      \langle x,\vec k_{\!\perp}; \lambda_{q},
        \lambda_{\bar q}  \vert \psi _i\rangle$ 
gives the probability amplitudes for finding in the $q\bar q$-state
a flavored quark with momentum fraction $x$, intrinsic transverse 
momentum $\vec k_{\!\perp}$ and helicity $\lambda_{q}$,
and correspondingly an anti-quark with $1-x$, $-\vec k_{\!\perp}$
and $\lambda_{\bar q}$.  
Both the mass and the wave-functions are boost-invariant. 

\section{Perturbation theory in medium}
\label{sec:b}

Here seems to be a problem:
For calculating $G_3$ one needs 
$G_6$, $G_5$ and $G_4$, see Eq.(\ref{eq:610}), 
for calculating $G_6$ one needs 
$G_{10}$, $G_9$ and $G_7$, see Eq.(\ref{eq:620}), 
and so on. 
This property corresponds to some extent the infinite series of
the Tamm-Dancoff approach, see Eq.(\ref{eq:m24}).
But the method of iterated resolvents resumes them 
systematically without double counting.
In the next section will be shown how the hierarchy can
be broken in a rather effective way.
That final step will be comparatively simple if one has analyzed 
the propagators for the sectors with one $q\bar q$-pair 
and arbitrarily many gluons, as follows next.

\begin{figure} 
 \resizebox{0.48\textwidth}{!}{%
 \includegraphics{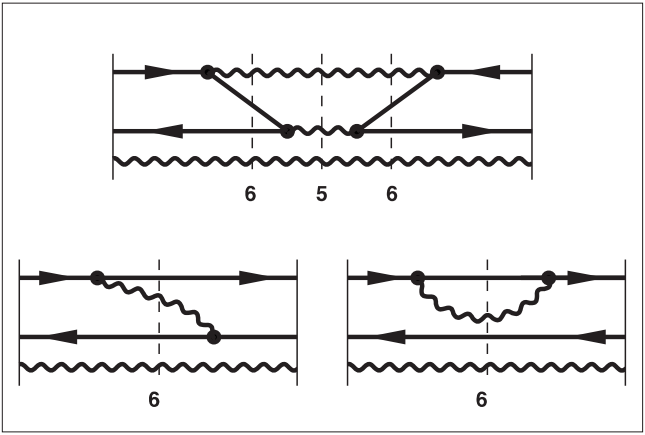}
}\caption{
    The three graphs of the  spectator interaction 
    in the $q\bar q\,g$-space. 
    Note the role of the gluon  as a spec\-tator.
}\label{fig:6_2} 
\end{figure}

\begin{figure} 
 \resizebox{0.48\textwidth}{!}{%
 \includegraphics{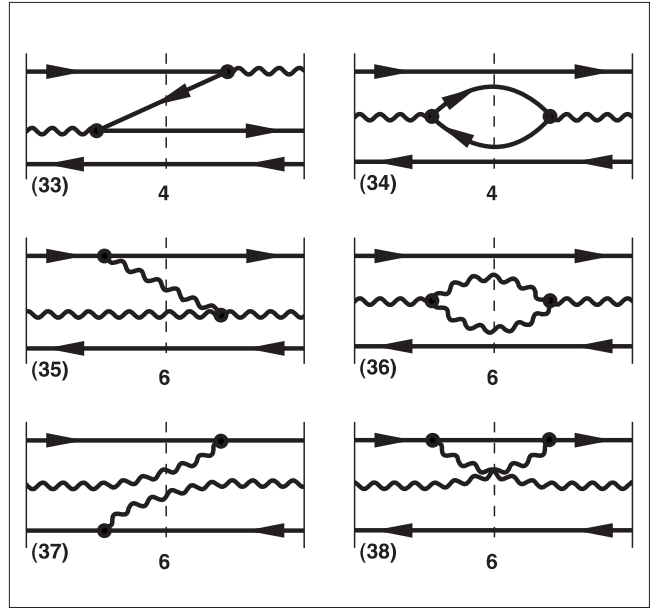}
}\caption{
    Some six graphs of the  participant interaction 
    in the $q\bar q\,g$-space. 
}\label{fig:6_3}
\end{figure}

Consider first the case with one gluon as given 
by Eq.(\ref{eq:620}).
The corresponding diagrams can be grouped into two 
topologically distinct classes, displayed 
in Figs.~\ref{fig:6_2}  and \ref{fig:6_3}. 
Adding one free gluon to the diagrams in Fig.~\ref{fig:6_1} 
produces the diagrams in Fig.~\ref{fig:6_2}.
The gluon does not change quantum numbers under impact
of the interaction and acts like a spectator.  Therefore, 
the graphs in Fig.~\ref{fig:6_2} will be referred to as the  
`spectator interaction' $\overline U _3$. 
In the graphs of Fig.~\ref{fig:6_3} the gluons are scattered
by the interaction, and correspondingly these graphs will be 
referred to as the `participant interaction' $\widetilde U _3$.  
Thus, $U_3 =  \overline U _3 + \widetilde U _3$.

Next consider the propagator $G _6$ in the $q\bar q\,gg$-space.
By drawing all possible diagrams of $U _6$, 
as given in Eq.(\ref{eq:621}), one realizes quickly that
most of them topologically belong to one of the
two classes which one obtains by adding a free gluon
to the diagrams in Figs.~\ref{fig:6_2} and ~\ref{fig:6_3}.
The effective interaction $V G_{10} V$, however, generates also 
an interaction between the two Fock-space gluons in
$q\bar q\,gg$, by an effective one-gluon exchange.
Potentially, these diagrams contribute to a $gg$-bound-state 
(a glue-ball), very much like the gluon-exchange between the 
quark and the antiquark contributes to a $q\bar q$-bound-state.
By definition, these glue-ball diagrams will be
included in the spectator interaction $\overline U _6$.
Estimates on their size will be given elsewhere \cite{pau98}.

The separation into spectators and participants  
can be made in all quark-pair-gluon sector interactions:
\begin {equation}  
     U_n =  \overline U _n + \widetilde U _n
\ , \quad{\rm for}\quad  n= 3,6,10,15,\dots
\ .  \label {eq:626} \end {equation} 
More explicitly, the spectator interactions in the lowest 
sectors with one quark-pair become for example
\begin {eqnarray} 
     \overline U_3 
&=& V G _6 V \ + V G _6 V  G _5 V G _6 V 
\,,\label{eq:n50}\\  
     \overline U_6 
&=&  V G _ {10} V+ V G _{10} V  G _9 V G _ {10} V 
\,.\label{eq:n51}\end {eqnarray}
Correspondingly the participant interactions are
\begin {eqnarray} 
     \widetilde U_3 
&=& V G _4 V\ + V G _6 V 
\,,\\  
     \widetilde U_6 
&=&  V G _7 V \ + V G _ {10} V
\,.\end {eqnarray}
The same operators appear in both 
$\overline  U$ and $\widetilde U$. 

Since the Hamiltonian is additive in spectator and participant
interactions, $\overline U _n$ can be associated with 
\begin{eqnarray} 
\lefteqn{ 
     \overline G _n  =   {1\over \omega-T_n-\overline U_n }
\,, }\nonumber\\ & &
    \biggl( {\rm while\ } G _n\equiv{1\over \omega-H_n} 
     ={1\over \omega-T_n-\overline U_n-\widetilde U_n}\ \biggr)
\,. \label{eq:641}\end{eqnarray} 
The relation to the full resolvent 
$G_n=\overline G_n+\overline G_n\,\widetilde U_n\,G_n$
is exact.         
Equivalently, it is written as an infinite series
\begin{equation} 
     G _n  = \overline G _n
     + \overline G _n \widetilde U _n\, \overline G _n 	
     + \overline G _n \widetilde U _n\, \overline G _n 
                                     \widetilde U _n\, \overline G _n   + \dots 
\,.\end{equation}  
The difference to the perturbative Tamm-Dancoff series is that the 
`unperturbed propagator' $ G _0 (\omega) $ in Eq.(\ref{eq:m24}) 
refers to the system without interactions 
while here the `unperturbed propagator' $\overline G _n$ contains 
the interaction in the well defined form of $\overline U_n$.
One deals here with a {\em perturbation theory in medium}.

The different physics should be emphasized: 
The system is not scattered into other sectors, 
it stays in sector $n$. 
This is reflected in an operator identity,
\begin{eqnarray} 
  (\omega - \overline H _n )(1-\widetilde U _n \overline G _n ) =
  \omega - H _n 
,\label{eq:6530}\end{eqnarray} 
which is obtained straightforwardly from Eqs.(\ref{eq:626})  and 
(\ref{eq:641}).
The inverse gives $R_n^2 \,\overline G _n = G _n$, with
\begin  {equation} 
      R_n = \sqrt{{1\over 1 - \overline G _n \widetilde U _n} }
\,. \label{eq:6520}\end {equation} 
With the obvious identity 
$R_n\,\overline G _n = \overline G _n\, R_n^\dagger$, 
one ends up with $G _n = R_n\,\overline G _n \,R_n^\dagger$.
In all of the above effective interactions the 
$R_n$ are sandwiched between a quark-pair-glue resolvent 
$\overline G$ and the vertex $V$,
\begin  {equation} 
     V\,G _n \, V ^\dagger = 
     V\,R_n\,\overline G _n \,R_n^\dagger\, V ^\dagger
     = \overline V \,\overline G_n\,\overline V ^\dagger     
\,.\label {eq:652} \end {equation} 
Each block matrix $V$ is rectangular and multiplied with 
a square matrix $R_n $ according to
\begin{equation}
       \overline V = V R_n , \qquad {\rm or}\qquad  
       \overline V ^\dagger = R_n^\dagger V ^\dagger  
\,.\end{equation}
In the sequel we shall suppress again the dagger indicating
the hermitean conjugate matrix.
In Section~\ref{sec:4} will be shown that $R_n$ is essentially 
diagonal and independent of the spin, 
such that each vertex matrix element is multiplied 
with a number, actually with a number which depends 
on the momentum transfer $Q$ across the vertex. 
Equivalently one replaces the coupling constant $g$ 
by an {\em effective coupling constant} $\overline g$, {\it i.e.} 
\begin{equation}
       g \longrightarrow \overline g (Q) = g R_n(Q)
\,.\end{equation}
$R_n$ has thus the interpretation of a {\em vertex function}.

One can thus rewrite 
Eqs.(\ref{eq:610}), (\ref{eq:n50}) and (\ref{eq:n51}) 
in such a form that they are all essentially equal, {\it i.e.}
\begin {eqnarray} 
     \overline U_6 
&=&   \overline V \,\overline G _ {10} \overline V
                     + \overline V \,\overline G _{10} \overline V  
              G _9 \overline V \,\overline G _ {10} \overline V 
\ , \label {eq:662} \\
     \overline U_3 
&=&   \overline V \,\overline G _6 \overline V 
                    + \overline V \,\overline G _6 \overline V  
              G _5\overline V \,\overline G _6 \overline V 
\ , \label {eq:663} \\      
       U_1
&=&  \overline V \,\overline G _3 \overline V 
                    + \overline V \,\overline G _3 \overline V  
             G _2 \overline V \,\overline G _3 \overline V 
\,.\label{eq:664} \end {eqnarray} 
The effective sector Hamiltonians  
$\overline H _n = T_n + \overline U _n$ describes bound 
states of one $q\bar q$-pair with arbitrarily many gluons
and glue balls. Approximatively, one can relate them
to each other, see Sec.~\ref{sec:c}. 

The content of Sects.~\ref{sec:3} and \ref{sec:b} 
is exact but rather formal. 
To show its usefulness, rigor will be given up in the sequel 
in favor of transparency 
and the aim to obtain a simple and solvable equation. 
It should be emphasized here already that the
content of Sects.~\ref{sec:c} and \ref{sec:4}
will have to be substantiated in future work \cite{pau98}.

\section{The breaking of the propagator hierarchy}
\label{sec:c}

All effective sector Hamiltonians can be diagonalized 
on their own merit.
To shape notation, the first few eigenvalue equations 
are written down explicitly. 
In the $q\bar q$-space they read in Fock space representation 
\begin{eqnarray} 
    M_{1;i} ^2
    \,\langle q;\bar q\vert\psi_i^{(1)}\rangle =
    \sum _{q^\prime,\bar q^\prime}
    \langle q;\bar q\vert 
    H_1 
    \vert q^\prime;\bar q^\prime\rangle
    \,\langle q^\prime;\bar q^\prime
    \vert\psi_i^{(1)}\rangle 
.\label{eq:6.60}\end{eqnarray}
For simplicity, the eigenvalues are enumerated by 
$i=0,1,2,\dots,\infty$. In the sequel, the
lowest eigenvalue is supposed to satisfy the fix-point equation  
$\omega=M^2_{1,0}(\omega)$ and its numerical value
will be denoted by $M^2$.
Correspondingly, one has in the $gg$-space 
\begin{eqnarray} 
    M_{2;j} ^2 
    \,\langle g;\bar g\vert\psi_j^{(2)}\rangle =
    \sum _{g^\prime,\bar g^\prime} 
    \langle g;\bar g\vert 
    H_2 
    \vert g^\prime;\bar g^\prime\rangle
    \,\langle g^\prime;\bar g^\prime
    \vert\psi_j^{(2)} \rangle 
,\label{eq:6.61}\end{eqnarray}
and in the $q\bar q\,g$ space
\begin{eqnarray} 
\lefteqn{ 
    \overline M _{3;j} ^2 
    \,\langle q;\bar q;g\vert\psi_{j}^{(3)}\rangle 
}\nonumber\\ & & =
    \sum _{q^\prime,\bar q^\prime,g^\prime}
    \langle q;\bar q;g\vert
    \overline H_{3}
    \vert q^\prime;\bar q^\prime;g^\prime\rangle
    \,\langle q^\prime;\bar q^\prime;g^\prime
    \vert\psi_{j}^{(3)}\rangle 
.\label{eq:6.62}\end{eqnarray}
Knowing the complete sets of eigenfunctions, 
one can calculate the exact resolvents
in Fock space representation,  
as for example the resolvent in the two gluon sector 
\begin{eqnarray} 
\lefteqn{ 
    \langle g;\bar g\vert 
    G_2 (\omega)
    \vert g^\prime;\bar g^\prime\rangle 
}\nonumber\\ & & =
    \sum _{j}
    \langle g;\bar g\vert\psi_j^{(2)} \rangle
    {1\over \omega - M_{2;j} ^2 }
    \langle \psi_j^{(2)} \vert g^\prime;\bar g^\prime\rangle
.\end{eqnarray}
As often in many-body physics, one approximates resolvents by
assuming that all eigenvalues are degenerate with the lowest state,
which here is the glue ball mass $M^2_{g}$. 
One then applies closure 
\begin{eqnarray} 
    \langle g;\bar g\vert 
    G_2 
    \vert g^\prime;\bar g^\prime\rangle =
    {1\over M^2 - M_{g} ^2}\sum _{j}
    \langle g;\bar g\vert\psi_j\rangle
    \langle \psi_j \vert g^\prime;\bar g^\prime\rangle 
\end{eqnarray} 
in order to obtain a diagonal resolvent
\begin{eqnarray} 
    \langle g;\bar g\vert 
    G_2 
    \vert g^\prime;\bar g^\prime\rangle =
    {1\over M^2 - M_{g} ^2}
    \langle g;\bar g\vert g^\prime;\bar g^\prime\rangle
.\label{eq:6.67}\end{eqnarray} 

The propagator in the $q\bar q g$-space could be calculated by
the same procedure, but one can do better.
Since the gluon is a free particle which moves
relative to the $q\bar q$-bound state, 
the eigenfunction 
$\vert \psi_{j}^{(3)}\rangle=\vert \psi_{i}^{(1)}\rangle
   \otimes \vert \varphi_{s}\rangle$ 
is a product state.
The $q\bar q g$-eigenvalues 
\begin  {equation}
    \overline M_{3;j} ^2 \equiv \overline M_{3;i,s} ^2
    = {M_{1;i}^2 + \vec q_{\!\perp}^{\,2} \over (1-y)} 
    + {\vec q_{\!\perp}^{\,2} \over y} 
\,,\end {equation}
can therefore be expressed in terms of the 
$q\bar q$-eigenvalues. 
Every $q\bar q$-bound state is band head for a continuum 
of gluons. 
With the four-momenta 
$   q^\mu=(yP^+,y\vec P_{\!\perp}+\vec q_{\!\perp},q_g^-)$ 
one gets for the lowest bound state 
\begin  {equation}
     \omega - \overline M_{3;0,s} ^2 
     \equiv M^2 - \overline M_{3;0,s} ^2 
     = -{y^2 M^2 + \vec q_{\!\perp}^{\,2} \over y(1-y)} 
\,.\label{eq:n70}\end {equation}
Assuming a degenerate spectrum and performing closure, 
\begin{eqnarray}
   \langle q;\bar q;g \vert \overline G _3
   \vert q ^\prime;\bar q^\prime;g ^\prime \rangle =
   \overline G _3 (q;\bar q;g)
   \langle q;\bar q;g\vert q ^\prime;\bar q^\prime;g^\prime\rangle 
,\label{eq:6.69}\end{eqnarray} 
one gets the delta function 
$\langle q;\bar q;g\vert q^\prime;\bar q^\prime;g^\prime\rangle$
multipied with 
\begin{eqnarray}
   \overline G _3 (q;\bar q;g) = -{y\over Q^2}
,  \ Q^2 = (y^2 M ^2 +\vec q_{\!\perp}^{\,2}){1\over 1-y}
.\label{eq:6.68}\end{eqnarray}
Note that Eqs.(\ref{eq:6.67})  and (\ref{eq:6.68}) 
break {\em the hierarchy of the iterated resolvents}.
For calculating  the effective interaction in the 
$q\bar q$-space only the two resolvents
$\overline G_3$ and $G_2$ are needed. 
Both are written now in closed expressions.
The whole complication of having resolvents of resolvents 
is replaced by the problem of knowing the eigenvalues
$M$ and $M_{g}$ ahead of time. Good initial guesses
($M_0$) might suffice but can be improved iteratively 
with $M_n\rightarrow M_{n+1}$.

The notation in Eq.(\ref{eq:6.68}) is suggestive  
for  $Q^2$ being related to the single-particle 
four-momentum transfer along the quark line. 
The free propagator in the  $q\bar q g$-space
can be written \cite{bpp97}
\begin{eqnarray}
      G_{q\bar q g,{\rm free}} &=& 
      {1\over P^+(p^--p^{\prime -} -q^-)} = 
      {y\over (p-p^\prime)^2}
\,.\label{eq:m72}\end{eqnarray}
The single-particle notation refers to Fig.~\ref{fig:g_1}. 
For sufficiently small $y$ holds 
$(p-p^\prime)^2= -[y^2(2\overline m)^2 +\vec q_{\!\perp}^{\,2}]$.
If one substitutes $M\simeq 2\overline m$, which holds to 
rather good approximation, the momentum transfer in 
Eq.(\ref{eq:m72}) is the same as in Eq.(\ref{eq:6.68}). 
One concludes: 
{\em In the solution, the interacting particles propagate 
like free particles to a high degree of approximation;
they just acquire an effective mass}.

Most importantly, instead of having resolvents of 
resolvents, the hierarchy of iterated resolvents is broken. 
{\it Ex post}, this justifies the $\omega = \omega ^\star$ trick in the 
Tamm-Dancoff work \cite{kpw92,trp96}. 

One can restore the exact $q\bar q g$-propagator 
by addition and subtraction, which gives
\begin{eqnarray}
\lefteqn{ 
   \langle q;\bar q;g \vert \overline G _3
   \vert q ^\prime;\bar q^\prime;g ^\prime \rangle =
   \overline G _3 (q;\bar q;g)
}\nonumber\\ & & \times\left[
   \langle q;\bar q;g \vert q ^\prime;\bar q^\prime;g^\prime\rangle -
   \langle q;\bar q;g\vert A\vert q ^\prime;\bar q^\prime;g^\prime\rangle 
   \right] 
.\label{eq:n75}\end{eqnarray} 
In Eq.(\ref{eq:6.69}) the operator $A$ was neglected, with
\begin{eqnarray} 
   A = \sum _{i,s} 
   \vert\psi_{i,s}^{(3)}\rangle\,{y(M_{1;i}^2-M^2)
   \over  Q^2+y(M_{1;i}^2-M^2)}\, 
   \langle \psi_{i,s}^{(3)} \vert 
\,.\end{eqnarray}
Probably, this is the most drastic assumption in this work. 
It suppresses the appearance of the Lamb-shift.

Note that the above approximations 
can be {\em controlled e posteriori}:  
Once the eigenfunctions in the $q\bar q$-sector are available
numerically, one can check whether  the exact definition of the 
resolvent Eq.(\ref{eq:6.69}) is peaked like a $\delta$-function, 
with a residue predicted by Eq.(\ref{eq:6.68}). 

\section{Perturbative analysis of the vortices}
\label{sec:4}

\begin{figure} 
 \resizebox{0.48\textwidth}{!}{%
 \includegraphics{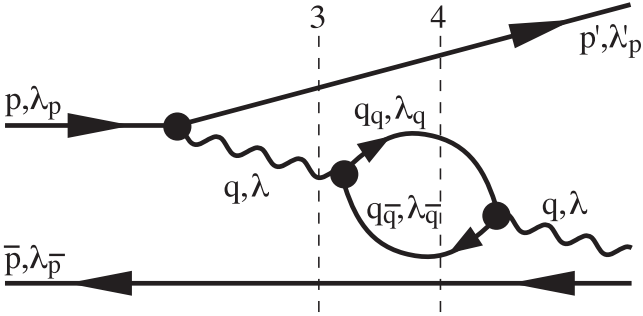}
}\caption{
The $q\bar q$ vacuum polarization graph.
}\label{fig:g_1} 
\end{figure}

\begin{figure} 
 \resizebox{0.48\textwidth}{!}{%
 \includegraphics{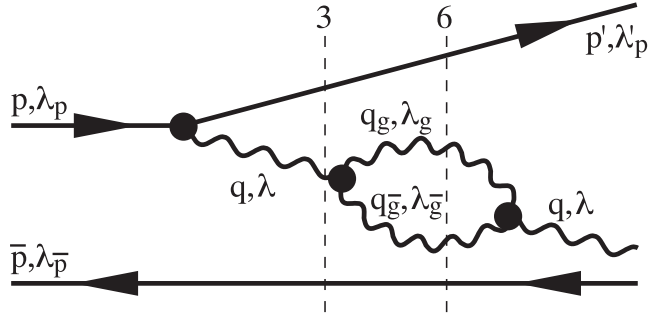}
}\caption{
The $gg$ vacuum polarization graph.
}\label{fig:g_2} 
\end{figure}

Having established that the propagator $\overline G_3$
can be approximated by the free propagator,
one can calculate the vertex function $R_3$ straightforwardly. 
According to Eq.(\ref{eq:6520}) one has
\begin  {equation}
      R_3 =  \sqrt{{1\over 1 - \overline G _3 \widetilde U _3}}
               =   1
               +  {1\over 2} \overline G _3 \widetilde U _3
               +  {3\over 8} \overline G _3 \widetilde U _3  \,
                           \overline G _3 \widetilde U _3 + \dots  
\,.\label{eq:m74}\end{equation} 
It is reasonable to restrict to the first non-trivial term in the
expansion
\begin{equation}
    \overline V = V R_3
    \simeq V+ {1\over2} \left( V\overline G_3V\overline G_4V+
    V \overline G _3 V\overline G _6 V \right)
\,.\label{eq:m73}\end{equation}
Since the contributions from the vertex functions $R_4$ and $R_6$
generate terms of higher orders in $g$, one must set $R_4=R_6=1$,
in order to be consistent.
A diagrammatic analysis of Eq.(\ref{eq:m73}) un-reveals quickly that 
it is the familiar `vertex correction',
which has been calculated in the front form 
by Thorn \cite{tho79} and by Perry \cite{phz93}. 
Thorn and later Perry, however, were motivated primarily
by asymptotic freedom `on the light-cone' 
and thus have emphasized the behaviour at large momentum 
transfer $Q\rightarrow\infty$.
In the present context one is interested in the opposite limit
$Q\rightarrow0$. In a bound state equation such
as Eq.(\ref{eq:445}) the region near the Coulomb singularity
$Q\sim0$ is very important.
Raufeisen \cite{rap97} has recalculated therefore all about 22 
diagrams of the vertex correction in light-cone perturbation theory,
see for example \cite{bpp97}.
The salient features of these calculations will be exposed here
at hand of the two vacuum polarization diagrams 
in Figs.~\ref{fig:g_1} and \ref{fig:g_2}. 

Next let us calculate the free propagator in the 
$q\bar q\,q\bar q$-sector, {\it i.e.}  
\begin{eqnarray} 
   \overline G _4 = 
   {1\over ({p}^- -{p'}^- -q_{q}^- -\bar q_{\bar q}^-)P^+}
.\end{eqnarray} 
Parameterizing in Fig.~\ref{fig:g_1} 
the intermediate particle momenta as 
\begin{eqnarray}
       q_{q}^\mu =
      \left(yzP^+, yz\vec P_{\!\perp} + z\vec q_{\!\perp}
      + \vec l_{\!\perp}, q_{q}^-\right),
      \bar q_{\bar q}^\mu =
\nonumber \\
      \left(y(1-z)P^+, y(1-z)\vec P_{\!\perp} + (1-z)\vec q_{\!\perp}
      - \vec l_{\!\perp},\bar q_{\bar q}^-\right)
,\end{eqnarray}
and using the four-momentum transfer along the quark line $Q$
as defined in Eq.(\ref{eq:6.68})
the propagator becomes 
\begin  {equation}
   \overline G _4 = 
   -{q_g^+\over P^+}{z(1-z)\over z(1-z)Q^2 
   + \vec l_{\!\perp}^{\,2}+m^2_f }
.\label{eq:freeG4}\end{equation}
Evaluating the spinor summations gives for $n_f$ flavors
\begin{eqnarray} 
\lefteqn{ 
   V\overline G_3V\overline G_4V = 
   - V\sum_{f=1}^{n_f} {\alpha\over 4\pi} \int \!dz\,dl_{\perp}^2
}\nonumber\\ & & 
   \Theta(z,l_{\perp}){1-2z(1-z) 
   -\bigg[{2m_f^2\over Q^2}\bigg]
   \over l_{\perp}^2+m_f^2+z(1-z)Q^2 }
.\end{eqnarray} 
Note the spin-independent factor to the matrix element $V$. 
The cut-off function $\Theta$ was defined in Eq.(\ref{eq:i3}).
The term in the square brackets are cancelled
by other diagrams and will be disregarded in the sequel.
Integrating over $l_{\perp}$ yields straightforwardly
\begin{eqnarray} 
\lefteqn{ 
      \overline G_3V\overline G_4V = - \sum_f {\alpha\over 4\pi} 
}\nonumber\\ & & 
      \int\limits_{\epsilon_f} ^{1-\epsilon_f} \!\!\!dz\,
      \bigg(1-2z(1-z) \bigg)
      \ \ln{ {\Lambda^2+4m_f^2+Q^2
      \over Q^2+{m_f^2\over z(1-z)} } }
\,.\end{eqnarray} 
The integral over $z$ is non-trivial. 
Since the logarithm is a very weak function
of its arguments, the term $m_f^2/z(1-z)$ is replaced 
by its maximum at $z={1\over2}$, thus  
\begin{equation}
      \overline G_3V\overline G_4V =-\sum_f {\alpha\over 6\pi} 
      \ln{\bigg( 1+{\Lambda^2\over 4m_f^2+Q^2}\bigg)}
\,.\end{equation}
The corresponding steps for the propagator in the 
$q\bar q\,gg$-sector (Fig.~\ref{fig:g_2}) give with
$\overline G_6 = 
\big[( {p}^- -{p'}^- - q_{g}^- - \bar q_{g}^- )P^+\big]^{-1}$
\begin{eqnarray}
   \overline G_6 = -{q_g^+\over P^+}{z(1-z)\over z(1-z)Q^2 +
   \vec l_{\!\perp}^{\,2}+m^2_g}
.\label{eq:freeG6}\end{eqnarray}
For the purpose of regularization
the gluon is endorsed with a small regulator mass $m_g$, 
see Sec.~\ref{sec:2}.

The spinor sums yield
\begin{eqnarray} 
\lefteqn{ 
   V\overline G_3V\overline G_6V = V\,{\alpha n_c \over 2\pi} 
   \int\! dz\,dl_{\perp}^2 }
   \nonumber\\ & & 
   \Theta(z,l_{\!\perp}) \,{2-z(1-z)
   \,-\bigg[{1\over z(1-z)}\bigg]
   \over l_{\perp}^2+m^2_g+z(1-z)Q^2 }
\,.\label{eq:rs82}\end{eqnarray} 
Dropping the term in the square bracket, and performing 
the same approximation as above gives
\begin{equation}
     \overline G_3V\overline G_6V ={11\alpha n_c \over 12\pi} 
      \ln{\bigg(1+ {\Lambda^2\over 4m^2_g+Q^2}\bigg)}
\,.\end{equation}
As a net result, the coupling constant $g$ has to be replaced
by the effective coupling constant $\overline g$, {\it i.e.} 
$g\rightarrow\overline g $ with
\begin{equation}
        \overline g = g R_3= 
        g \left(1+{1\over 2} \alpha b(Q) \right)
\,,\label{eq:m82}\end{equation}
see also \cite{rap97}. The function $b(Q)$ depends 
on the momentum transfer (and on the cutoff), {\it i.e.}
\begin{eqnarray}
\lefteqn{
   b(Q) = {11n_c \over 12\pi} 
   \ln{\bigg(1+{\Lambda^2\over 4m^2_g+Q^2}\bigg)} -
}\nonumber\\ & & 
   {1\over 6\pi} \sum_f 
   \ln{\bigg(1+{\Lambda^2\over 4m_f^2+Q^2}\bigg)}
\,\label{eq:oi63}\end{eqnarray}
and generates therefore a new interaction.
Finally, the effective quark masses become
according to the diagram $U_{1,2}$ in Fig.~\ref{fig:6_1}
\begin{equation}
      \overline m_f^2 =m_f^2 +m_f^2 {\alpha\over \pi}
      {n_c^2-1\over 2n_c}
      \ln{ {\Lambda^2\over m_{g}^2} }
\,.\label{eq:i64}\end{equation}
A similar diagram for the effective gluon mass gives
\begin{equation}
      \overline m_g^2 =m^2_g -{\alpha\over 4\pi}
      \sum_f  m_f^2\ \ln{\bigg(1+{\Lambda^2\over 4m_f^2}\bigg) }
\,.\label{eq:g2}\end{equation}
Both are obtained by light-cone perturbation 
theory \cite{bpp97} combined with vertex regularization.

The above considerations are perturbative, as mentioned.
What happens if one substitutes the free propagators
in Eqs.(\ref{eq:freeG4}) and (\ref{eq:freeG6}) by
the non-perturbative propagators
$\overline G_4= (\omega- \overline H_{q\bar q\,q\bar q})^{-1}$ and
$\overline G_6= (\omega- \overline H_{q\bar q\,gg})^{-1}$,
at least in an approximate fashion?~---
There are additional graphs.
In the fermion loop of vacuum polarization appear two 
graphs in addition to Fig.~\ref{fig:g_1}.
In one of them, a gluon is emitted and absorbed 
on the same quark line 
which changes the bare quark mass $m _f$
into the  physical quark mass $\overline m _f$.
In the other graph, the gluon is emitted from 
the quark and absorbed by the anti-quark which
represents an interaction. In consequence one has
a bound state with a physical mass scale $\mu_f$.
We replace therefore
\begin{equation}
   2m _f \Longrightarrow 2\overline m _f \Longrightarrow \mu _f
.\end{equation}
Similar considerations hold for the gluon loop
in Fig.~\ref{fig:g_2} and lead to the substitution
\begin{equation}
   2m _g \Longrightarrow 2\overline m _g \Longrightarrow \mu _g
.\end{equation}
Both $\mu_g$ and $\mu_f$ are interpreted as physical mass scales.
The physical gluon mass $\overline m _g$ vanishes of course 
due to gauge invariance. This is not in conflict
with {\it f.e.} Cornwall's suggestion of a finite effective 
gluon mass \cite{cor83} since one can define
$(\overline m _g)_{\rm eff}\equiv \mu_g/2$.
Estimates are given in Eq.(\ref{eq:estimates}), 
and discussed in Sec.~\ref{sec:6}.
In conclusion, for sufficiently large $\Lambda$, 
one replaces Eq.(\ref{eq:oi63}) by
\begin{equation}
   b(Q) = {11n_c \over 12\pi} 
   \ln{\bigg({\Lambda^2\over \mu^2_g+Q^2}\bigg)} - 
   {1\over 6\pi} \sum_f 
   \ln{\bigg({\Lambda^2\over \mu_f^2+Q^2}\bigg)}
.\label{eq:i63}\end{equation}
For $Q\gg (\mu_g,\mu_f)$ this becomes
\begin{equation}
   b(Q) = b_0 \ln{\bigg({\Lambda^2\over Q^2}\bigg)}
   ,\quad{\rm with}\qquad
   b_0 = {33 -2n_f\over 12\pi} 
,\label{eq:ni63}\end{equation}
for $n_c=3$.
It coincides with the perturbative expressions 
of Thorn and of Perry.
The notation is inspired by the relation to the familiar 
beta-function \cite{wei95}.

\section{Renormalization}
\label{sec:5}

Now, all the pieces are together which are needed for a further
discussion of the effective interaction as defined in  Eq.(\ref{eq:445}).
In the present first assault the flavor changing interaction $U_a$ is 
disregarded. In the remainder $U$, 
the instantaneous interaction is restored by  $\overline U_{q\bar q}
= W_{q\bar q} + \overline V\,\overline G_{q\bar q\,g}\overline V$,
with $\overline G_{q\bar q\,g}$ given by Eq.(\ref{eq:6.68}).
Both terms contain a non-integrable singularity
which however cancel each other such that only the integrable 
Coulomb singularity $Q^{-2}$ remains, 
see for example \cite{bpp97}. 
When substituting in both of them  $g\rightarrow \overline g $ 
one gets from Eq.(\ref{eq:445}) the final one-body equation:
\begin{eqnarray} 
\lefteqn{ 
    M^2\langle x,\vec k_{\!\perp}; \lambda_{q},
    \lambda_{\bar q}  \vert \psi\rangle =
}\nonumber\\ & & \left[ 
    {\overline m^{\,2}_{q} + \vec k_{\!\perp}^{\,2}\over x } +
    {\overline m^{\,2}_{\bar  q} + \vec k_{\!\perp}^{\,2}\over 1-x }   
    \right]\langle x,\vec k_{\!\perp}; \lambda_{q},
    \lambda_{\bar q}  \vert \psi\rangle 
\nonumber\\ & & 
     - {1\over 3\pi^2}
    \sum _{ \lambda_q^\prime,\lambda_{\bar q}^\prime}
    \!\int\!dx^\prime d^2 \vec k_{\!\perp}^\prime
    \,\Theta(x^\prime,k_{\perp}^\prime) 
\nonumber\\ & & 
    {\overline\alpha(Q) \over Q  ^2} \,\langle
    \lambda_q,\lambda_{\bar q}\vert S(Q)\vert 
    \lambda_q^\prime,\lambda_{\bar q}^\prime\rangle
    \,\langle x^\prime,\vec k_{\!\perp}^\prime; 
    \lambda_q^\prime,\lambda_{\bar q}^\prime  
    \vert \psi\rangle 
.\label{eq:i65}\end {eqnarray}
The effective coupling constant 
$\overline \alpha (Q)= \overline g ^2(Q)/4\pi$
has thus far been given in Eq.(\ref{eq:m82}); its 
renormalized version will be given in Eq.(\ref{eq:m97}), below.
The cut-off function $\Theta$ sets the domain of integration and 
was defined in Eq.(\ref{eq:i3}).
The spinor factor $S(Q)$ represents the familiar current-current 
coupling which describes all fine and hyperfine interactions
\begin{eqnarray} 
\lefteqn{ 
    \langle
    \lambda_{q},\lambda_{\bar q}\vert S(Q)\vert 
    \lambda_{q}^\prime,\lambda_{\bar q}^\prime\rangle =
}\nonumber\\ & & 
    {\left[ \overline u (k_q,\lambda_q)\gamma^\mu
    u(k_q^\prime,\lambda_q^\prime)\right] 
    \over \sqrt{x x^\prime} } \,
    {\left[ \overline u (k_{\bar q},\lambda_{\bar q}) 
    \gamma_\mu
    u(k_{\bar q}^\prime,\lambda_{\bar q}^\prime)\right] 
    \over \sqrt{(1 - x) (1- x ^\prime)} } 
.\end{eqnarray} 
Actually, the precise analysis of the formalism requires 
to use the arithmetic mean from each vertex, {\it i.e.}
\begin{equation} 
    {\overline\alpha(Q) \over Q  ^2} \Longrightarrow  {1\over 2} 
    \sqrt{\overline\alpha(Q_{q}) \overline\alpha(Q_{\bar q}) }
    \left({1\over Q_{q}^2} +
    {1\over Q_{\bar q}^2} \right)
\,,\end{equation} 
where $Q^2_{q}=-(p_q-p_q^\prime)^2$ and 
$Q^2_{\bar q}=-(p_{\bar q}-p_{\bar q}^\prime)^2$ are the 
momentum transfers along the quark and the antiquark line, 
respectively. 
Close to the Coulomb singularity, however, this does not matter
since both are approximately equal:  
$Q^2_{q}\simeq Q^2_{\bar q} \simeq 
        (m_q +m_{\bar q})^2(x-x^\prime)^2 +
        (\vec k_{\!\perp}-\vec k_{\!\perp}^\prime)^2$.

Let us return to Eq.(\ref{eq:m82}).
The analysis there was incomplete and limited to the 
lowest perturbative order rather than to resume the series
to all orders in $\alpha$ as required by the definition of the 
vertex function, Eq.(\ref{eq:m74}). 
One can calculate the second order
diagram if one neglects all genuine two-loop contributions.
More generally, if one neglects all $n$-loop contributions
one can calculate the term of $n$-th order by
\begin{equation}
     \left(\overline G_3V\overline G_4V+
     \overline G _3 V\overline G _6 V 
     \right)^n \simeq \left(\alpha b(Q) \right)^n
,\end{equation}
and resume the series formally to all orders, {\it i.e.}
\begin{equation}
   \overline g(Q) = {g \over \sqrt{1 - \alpha b(Q)}}
   ,\ {\rm thus}\quad
   \overline \alpha(Q) = {\alpha\over 1 - \alpha b(Q) }      
.\label{eq:m96}\end{equation}
The restriction to the one-loop level is a customary procedure 
and not coupled directly to the size of $\alpha$. 
More important is the point that the series converges only if
\begin{equation}
      \alpha b(Q)\leq 1
\,,\label{eq:m90}\end{equation}
for all $Q$ and $\alpha$.
Let us therefore discuss shortly to which extent 
Eq.(\ref{eq:m90}) holds true. 
The possibility of $\alpha\ll 1$ is disregarded, 
since considerations should not be limited 
by perturbative considerations.
According to Eq.(\ref{eq:i63}), $b(Q)$ is largest for $Q=0$.
Defining a weighted average $\mu$ by
\begin{equation}
      b_0 \ln{\bigg({\Lambda^2\over \mu^2}\bigg)}
      \equiv {33\over 12\pi} 
      \ln{\bigg({\Lambda^2\over \mu^2_g}\bigg)}
      -{1\over 6\pi} \sum_f 
      \ln{\bigg({\Lambda^2\over \mu_f^2}\bigg)}
,\label{eq:average}\end{equation} 
gives $\mu^2 \ge \Lambda^2 e^{-{1\over\alpha b_0}}$
from Eq.(\ref{eq:m90}).
This lower limit is not necessarily small. An estimate
appropriate for charmonium ($M_c$) 
(with $\Lambda\simeq M_c$ and 
$\alpha b_0 \sim 0.2$) rather gives 
$\mu\ge 245 {\,\rm MeV}$. 
Denote the six experimental masses for the  
neutral scalar mesons with $M_f$ and the estimated glue-ball 
mass with $M_g\sim 1.4$ GeV. 
The scales $\mu_f$ and $\mu_g$ are identified with them
for simplicity. One gets from Eq.(\ref{eq:average}) 
\begin{equation}
   \mu_g=M_g,\ \mu_f=M_f,
   \quad {\rm thus}\quad \mu = 0.91\ {\rm GeV}
,\label{eq:estimates}\end{equation} 
independent of $\Lambda$ and consistent with the above limit. 

This is as far as one can go with the regulated theory.
Next let us address to the renormalization of the theory,
particularly to the renormalization group equations.
The eigenvalues of the light-cone Hamiltonian $H_{LC}$ 
are given by Eq.(\ref{eq:i65}). Potentially, they depend on 
$\Lambda$ through three sources: 
(1) the physical flavor masses $\overline m_f$;
(2) the $\Theta$-function representing the domain of integration;  
(3) the effective coupling constant $\overline \alpha(Q) $. 
Only the last one is relevant, since the flavor masses 
are renormalization group invariants with 
$d\overline m_f/d\Lambda=0$, and since 
(at least in a confining regime)  the wavefunctions decay 
sufficiently fast to serve as a natural cut-off.
It is thus reasonable to replace Eq.(\ref{eq:i69}) by
\begin{equation} 
    {d\over d\Lambda}\,\overline \alpha(\alpha;\Lambda) = 0
\,.\label{eq:m88}\end{equation}
It can be satisfied by a cut-off dependent bare coupling 
constant $g=g(\Lambda)$, thus by
\begin{equation}
      {d\overline \alpha\over d\Lambda} = 
      {1\over \displaystyle \left(1 - \alpha b(Q)\right)^2} \left(
      {d\alpha\over d\Lambda} + \alpha^2 {db\over d\Lambda}
       \right) = 0
\,.\end{equation}
Since $db/d\Lambda$ is completely independent of the mass scales
$\mu_g$ and $\mu_f$ appearing in Eq.(\ref{eq:i63}), {\it i.e.}
\begin{equation}
      {db(Q)\over d\Lambda} = 2{b_0\over \Lambda}
\,,\end{equation}
one gets the differential equation
${d\alpha/ d\Lambda} = - 2\alpha^2 b_0/\Lambda$,
as usual, including its solution \cite{wei95}
\begin{equation}
      \alpha_\Lambda = {\alpha_0 \over  
      1 + \alpha_0 b_0  \ln{{\Lambda^2\over\Lambda^2_0}}} =
      {1\over   b_0 \ln{{\Lambda^2\over\Lambda_{QCD}^2}}}
\,.\label{eq:m98}\end{equation}
The renormalization point $\alpha_0$ at the scale 
$\Lambda_0$ is arbitrary. 
The two parameters $(\alpha_0,\Lambda_0)$ 
can therefore be compounded into a single one, into 
{\em the QCD-scale} $\Lambda_{QCD}\equiv\kappa$ which is defined by
$2\alpha_0 b_0  \ln{\Lambda_0/\Lambda_{QCD}}=1$.
Substituting Eq.(\ref{eq:m98}) back into into Eq.(\ref{eq:m96}) 
gives 
\[
   \overline \alpha(Q) =
   {1\over \displaystyle b_0 
   \ln{{\Lambda^2\over \displaystyle \kappa^2}} 
   \bigg(1 - {b(Q)\over \displaystyle b_0 
   \ln{{\Lambda^2\over \displaystyle \kappa^2}} }\bigg) }=
   {1\over \displaystyle b_0 
   \ln{{\Lambda^2\over \displaystyle \kappa^2}} - b(Q)}
,\]
and thus 
\begin{equation}
      \overline \alpha(Q) 
      = {12\pi\over \displaystyle 
      33      \ln{\bigg({\mu^2_g+Q^2\over\kappa^2}\bigg)} - 
      2\sum_f \ln{\bigg({\mu^2_f
      +Q^2\over\kappa^2}\bigg)} }
,\label{eq:m97}\end{equation}
All $\ln{(\Lambda)}$ terms cancel exactly without smallness assumption.
The cancelation is directly related to the 
minus  sign in Eq.(\ref{eq:m96}), which in turn is related to the 
re-summation of the series {\em to all orders in} $\alpha$.
Using Eq.(\ref{eq:m82}), the cancelation would 
{\em hold only perturbatively}, for sufficiently small $\alpha$. 

Since $\overline \alpha$ becomes independent of $\Lambda$,
one can go to the limit $\Lambda\rightarrow\infty$.
In line with  the modern interpretation of the renormalization 
group one can integrate Eq.(\ref{eq:i65}) 
unrestrictedly over all $Q$.
Note the subtle difference between the 
`running coupling constant' $g _\Lambda$
and the `effective coupling constant' $\overline g(Q)$.
The two are confused often in the literature.
The Lagrangian does not know about the momentum transfer.


A final word on the complete wavefunction $\vert\Psi\rangle$.
The solutions of the integral equation Eq.(\ref{eq:i65}) represent
the normalized projections 
of the full eigenfunction $\vert\Psi\rangle$ onto the Fock states
$ \vert q;\bar q\rangle 
   = b^\dagger _q d^\dagger_{\bar q} \vert vac \rangle $, {\it i.e.}
$\psi^{(1)}=\langle1\vert\Psi\rangle$. 
Unlike in other methods 
this knowledge is sufficient in the present method 
to retrieve all other Fock-space components 
$\langle n\vert\Psi\rangle$ without
solving another eigenvalue problem.  
As an example, let us work that out explicitly, 
by asking for the probability amplitude to find a  
$\vert gg\rangle$- and a  $\vert q\bar q\,g\rangle$-state
in a full eigenstate $\vert\Psi\rangle$.

The key is the upwards recursion relation Eq.(\ref{eq:410}).
The first two equations of the recursive set in Eq.(\ref{eq:410}) are
\begin{eqnarray} 
   \langle 2 \vert\Psi\rangle &=& 
   G _ 2 \langle2\vert H_2\vert1\rangle 
   \langle 1 \vert \Psi  \rangle 
,\label{eq:448}\\  
   \langle 3 \vert\Psi\rangle  &=& 
   G _ 3 \langle 3 \vert H _3 \vert 1 \rangle 
   \langle 1 \vert \Psi  \rangle +
   G _ 3 \langle 3 \vert H _3 \vert 2 \rangle 
   \langle 2 \vert \Psi  \rangle 
\,.\end{eqnarray}
The sector Hamiltonians $H _n$ have to be substituted from
Eqs.(\ref{eq:620}) and (\ref{eq:643}).
In taking block matrix elements of them, the formal expressions
are simplified considerably since many of the Hamiltonian blocks 
in Table~\ref{tab:3} are zero.  
One thus gets simply
$\langle2\vert H _2\vert1\rangle=\langle2\vert VG_3V\vert1\rangle$
and therefore
$\langle2\vert\Psi\rangle=G_2VG _3V \langle1\vert\Psi \rangle$.
Substituted into Eq.(\ref{eq:448}) this gives 
$\langle 3 \vert \Psi \rangle  = G _ 3 V\langle 1 \vert\Psi \rangle +
    G _ 3 VG _ 2 VG _3 V \langle1\vert\Psi\rangle$. 
These findings are summarized more readable
\begin {eqnarray} 
\lefteqn{ 
    \langle gg \vert \Psi \rangle  = 
    \langle gg 
    \vert G _ {gg} VG _{q\bar q\,g}V\:\vert\psi_{q\bar q}\rangle  
,}\\  & & 
    \langle q\bar q\,g \vert\Psi \rangle  =
    \langle q\bar q\,g
    \vert G _ {q\bar q\,g} V\:\vert\psi _{q\bar q}\rangle 
\nonumber\\ & &  + 
    \langle q\bar q\,g 
    \vert G _ {q\bar q\,g} VG _ {gg} VG _{q\bar q\,g} V 
    \:\vert\psi_{q\bar q}\rangle 
.\end{eqnarray}
Correspondingly, one is able to find the Fock-space amplitude
in the higher sectors with remarkably little effort.
Of course, one has to readjust the overall normalization of the 
wavefunction. 
It should be emphasized that the finite number of 
terms is in strong contrast to the infinite 
perturbative series. Iterated resolvents resume the 
series to all orders in closed form. 
It also should be emphasized that the same approximations as 
discussed above must be made for reasons of consistency, 
particularly the effective coupling constant $\overline g(Q)$
must be used. 

\section {Summary and discussion} 
\label {sec:6}

The present work is based on Lagrangian gauge field theory
for SU(N) and on its canonical front-form Hamiltonian 
in the light-cone gauge $A^+=0 $. 
The notational background is laid down in 
Sects.~\ref{sec:2} and \ref{sec:a}.
All zero modes are disregarded, 
and in consequence the vacuum is trivial.
Imposing periodic boundary conditions 
the Hamiltonian is converted to a matrix. 
The matrix is finite by means of Fock-space regularization.
All possible divergences are controlled by vertex regularization. 
Being confronted with the diagonalization of a finite but large 
matrix, the problem is mapped on a smaller space as was done
first by Tamm and by Dancoff in their theory of effective interactions. 
Binding effects arise then due to
virtual scattering into higher Fock-space sectors.
The apparent difficulties with this approach are related to 
the infinite perturbative series one is forced to work with in practice.
A re-analysis shows that the main idea can be maintained 
if one introduces a hierarchy of effective interactions, 
in each sector of the Hilbert space. 
The result is an iterative procedure, 
called the method of iterated resolvents.
Each resolvent is the inverse of an effective sector Hamiltonian 
which in turn is a functional of resolvents in higher sectors. 
For the model case of a finite matrix (DLCQ) 
the method of iterated resolvents can be realized and 
explicitly checked by a finite number of successive 
matrix inversions and multiplications. 
But even for the continuum it is a well
defined and exact procedure. 
The infinite perturbative series of the Tamm-Dancoff approach 
and all their many-body aspects are then
replaced by a finite number of terms.
The effective interaction between a quark and an antiquark
turns out to have only two contributions:
The flavor conserving interaction $U$ and flavor changing 
interaction $U_a$. Their diagrammatic representations
look like second order diagrams of perturbation theory, but 
represent a re-summation of perturbative graphs to all orders. 
Particularly $U$ bears great similarity with a perturbative
one-gluon-exchange interaction. 
As part of the approach the complete wavefunction can be 
reconstructed, by evaluating
one component after the other in a well defined procedure.
Once this is achieved one can relax periodic boundary 
conditions and return to the continuum limit.
The result is a complete and exact theory of the effective 
interaction between a quark and an antiquark. 
It can be interpreted as the genuine interaction 
in a constituent quark model. 
The body of this work is found in 
Sects.~\ref{sec:3} and \ref{sec:b}.
 
In Sects.~\ref{sec:c} and \ref{sec:4} essentially
four assumptions are made for the sake of transparency 
which shall be summarized in brief.~---
The neglect of the operator $A$ in Eq.(\ref{eq:n75}) is probably 
the most drastic assumption.
It prevents a straightforward calculation of the Lamb shift
and reminds to proposition 13 in 
Feynman's famous article \cite{fey49}. 
It is here where statistical considerations \cite {vwz85}
can perhaps be implemented in the future. 
The next severe approximation resides in Eq.(\ref{eq:6.68}), 
where the eigenvalue $M$ appears in the propagator.
If one substitutes $M \simeq 2\overline m$,
the propagator in the $q\bar qg$ space behaves 
like the propagator of a free particle with an effective mass 
$\overline m$ (rather than with the bare mass $m$), 
to a high degree of approximation.
This step is important since it `breaks the hierarchy':
The propagator $\overline G_{q\bar q g}$ 
need not be obtained from an iteration procedure.
In a similar way one gets a simple approximation for $G_{gg}$. 
Only these two propagators occur explicitly in the expression
for the effective interaction.
By the nature of the approximation both of them become
independent of the starting point energy $\omega$.
The third and fourth assumption resides in the vertex function,
which contains all many-body effects to arbitrary order 
of the coupling constant. The vertex function was evaluated 
in Eq.(\ref{eq:m96}) up to the one-loop level, as done often 
in applications of gauge theory.~--
The following point should be emphasized:
Simplifying assumptions are made here only after having 
found the general structure. 
They are therefore unlikely to violate fundamental symmetries 
like gauge and Lorentz-invariance. 
Usually one proceeds in the reverse order
\cite{wwh94,gla97,brp95,bpw97,jop96b,kpw92,trp96,pab85a,lew96,guw97}: 
One first truncates and then develops the formalism.
  
Taking all together in Sect.~\ref{sec:5} one arrives at
a comparatively simple integral equation in the variables of
a single quark, the one-body equation (\ref{eq:i65}).
Its kernel contains the effective coupling constant
$\overline \alpha$. It is defined in Eq.(\ref{eq:m97}) and  
accumulates approximatively all many-body aspects. 
It depends on the four-momentum transfer $Q$ 
along the quark line and on the QCD scale 
$\Lambda_{QCD}\equiv\kappa$. The latter arises by renormalization
and must be determined by experiment.
Beyond that, $\overline \alpha(Q)$ has a rich parametric
structure depending on one gluonic and six fermionic mass scales,
$\mu_g$ and $\mu_f$, respectively.
They are in principle calculable 
\cite{pau98}, but as explained
in the context of Eq.(\ref{eq:6.68}), the present formalism needs 
an initial input guess which later can be improved 
iteratively and self-consistently. 
At present $\mu_g$ and $\mu_f$ are taken as external parameters.  
The values quoted in Eq.(\ref{eq:estimates}) 
look like reasonable first guesses. 

But one can view $\mu_g$ and $\mu_f$ also as external parameters,
which are fixed subject to describe experiments or other
theories. This opens a broad avenue for QCD-based 
phenomenological applications. 
One knows where the assumptions have been made in the present 
formalism, and can relax them in subsequent work.

For example, one can choose $\mu_g=\mu_f=\kappa$ to get 
$\overline \alpha(Q)^{-1}= b_0\ln{(1+Q^2/\kappa^2)}$. 
The numerical value of $\kappa = 0.19$ GeV 
gives a reasonable fit to the
empirical masses of the scalar and vector mesons \cite{pam95}.
This form was introduced by long ago Richardson \cite{Ric79} 
to interpolate between asymptotic freedom and infrared slavery, 
see also \cite{cor83}.
It generates a Coulomb plus a linear potential in configuration 
space, and thus produces confinement. 
``There is a pleasure in recognizing old things from 
a new point of view'' \cite {fey48}.
Unfortunately the Richardson coupling has the unphysical aspect 
of tending to infinity for vanishing four-momentum transfer.
But the full expression Eq.(\ref{eq:m97}) is finite for $Q=0$
since $\overline \alpha(0)^{-1}= b_0\ln{(\mu^2/\kappa^2)}$,   
with a numerical value $\mu=0.91$ GeV 
as given in Eq.(\ref{eq:estimates}). 
The parametrization
\begin{equation}
   \overline \alpha(Q)= {1\over b_0\ln{{\mu^2 + Q^2\over \kappa^2}} }
\label{eq:f_short}\end{equation}
is simple and suggestive as an approximation. It has been used 
repeatedly in the past, see {\it f.e.} \cite{cor83}, 
as reviewed in \cite{bjp98}.
Brodsky {\it et al} \cite{bjp98} have fixed the parameters by a fit
to the non-relativistic heavy-quark lattice data \cite{nrq95} 
and get $\kappa=0.16$ GeV and $\mu=0.87$ GeV.
The (perhaps accidental) coincidence with the 
above quoted numbers is amazing.

It would be more than interesting to solve the integral 
equation (\ref{eq:i65}), or approximations thereof, with various
parametrization of $\overline \alpha(Q)$, or at least calculate
the potential in configuration space. 
The work is under way but for obvious reasons must be
dealt with separately \cite{pau98}.
The investigation of chiral symmetry (breaking) must also be 
postponed for the future. In the present analysis it
cannot be discussed since the flavor changing interaction 
$U_a$ was disregarded waiting to be calculated.
It also should be interesting to apply the method of iterated
resolvents to conventional many-body problems, and 
to re-analyze QED.
  
To summarize the present work in short one can state that the 
effective potential between the constituent
quarks in a meson has been derived from the bare Lagrangian
for (actually any) SU(N) gauge field theory.
It is the first time that this was possible within a 
self-consistent
treatment. The approach is based a novel technology 
and has some virtues, among them: 
\\ $\bullet$\ %
The minimum number of physical degrees of freedom 
are used because of the light-cone gauge; 
\\ $\bullet$\ %
All Lagrangian symmetries are preserved;
\\ $\bullet$\ %
Renormalization is carried out explicitly;
\\ $\bullet$\ %
Fermions are treated exactly; thus no fermion doubling;
\\ $\bullet$\ %
The final one-body equation is solvable. 
\\ A lot of work remains to be done.

\begin{acknowledgement}
It is a pleasure to thank 
Antonio Bassetto,
Stan Brodsky,
Lloyd Hollenberg,
Paul Hoyer,
Steve Pinsky and 
Uwe Trittmann 
who have contributed to this work in one way or an other.
I thank them for their advice, their repeated 
suggestions, their nagging insistence, and last not least, 
for their patience with me in many fruitful discussions.
\end{acknowledgement}

\end{document}